\documentclass[10pt]{iopart}
\usepackage{iopams}
\usepackage{psfrag}
\usepackage{amssymb}
\usepackage{mathrsfs}
\usepackage{subfigure}

\usepackage{graphics}
\usepackage{epsfig}
\usepackage{psfrag}
\usepackage{subfigure}
\usepackage{citesort}
\usepackage{amsthm}

\renewcommand{\theequation}{\arabic{equation}}
\newtheorem{thm}{Theorem}

\newtheorem{conj}[thm]{Conjecture}

\begin{document}
\title{Knotting probabilities after a local strand passage in unknotted self-avoiding polygons}
\author{M. L. Szafron
and
C. E. Soteros
}

\address{
Department of Mathematics and Statistics,
University of Saskatchewan, Saskatoon  SK, Canada S7N 5E6 \\
\vspace{.15in}
}
\eads{\mailto{szafron@math.usask.ca},  \mailto{soteros@math.usask.ca}}

\begin{abstract}
We investigate, both theoretically and numerically, the knotting probability after a local strand passage is performed in an unknotted self-avoiding polygon on the simple cubic lattice.  In the polygons studied,
it is assumed that two polygon segments have already been brought close together for the purpose of performing a strand passage.  This restricts the polygons considered to those that contain a specific pattern called $\Theta$ at a fixed location; an unknotted polygon containing $\Theta$ is called a $\Theta$-SAP. It is proved that the number of $n$-edge $\Theta$-SAPs  grows exponentially (with $n$) at the same rate as the total number of $n$-edge unknotted self-avoiding polygons (those with no prespecified strand passage structure).
Furthermore, it is proved that the same holds for subsets of $n$-edge $\Theta$-SAPs that yield a specific after-strand-passage knot-type.  Thus the probability of a given after-strand-passage knot-type does not grow (or decay) exponentially with $n$.  Instead, it is conjectured that these after-strand-passage knot probabilities approach, as $n$ goes to infinity,  knot-type dependent amplitude ratios lying strictly between 0 and 1.  This conjecture is supported by numerical evidence from Monte Carlo data generated using a composite (aka multiple) Markov Chain Monte Carlo BFACF algorithm developed to study $\Theta$-SAPs.  A new maximum likelihood method is used
to estimate the critical exponents relevant to this conjecture.    We also obtain strong numerical evidence that the
after-strand-passage knotting probability depends on the local structure around the strand passage site.  
If the local structure and the crossing-sign at the strand passage site are considered, then 
we observe that the  more ``compact'' the local structure, the less likely the after-strand-passage polygon is to be knotted. This trend for compactness versus knotting probability is consistent with results obtained for other strand-passage models, however, we are the first to note the influence of the crossing-sign information.
We use two measures of ``compactness'': one involves the size of a smallest polygon that contains the structure and the other is in terms of an ``opening'' angle.  The opening angle definition is consistent with one that is measurable from single molecule DNA experiments.  
The theoretical and numerical approaches presented here are more broadly applicable to other
self-avoiding polygon models.

\end{abstract}
\section{Introduction}
Experimental evidence indicates that enzymes (type II topoisomerases) act locally in DNA to perform a strand passage  (two strands of the DNA which are close together pass through one another) in order to disentangle the DNA so that normal cellular processes
can proceed \cite{bm05}. Given that these enzymes only act locally, the DNA experiments of Rybenkov \etal \cite{ruvc97} show that type II topoisomerases reduce knotting (a global property) in DNA remarkably efficiently (the steady-state fraction of knots was found to be as much as 80 times lower than at equilibrium).   Experimentalists have not yet completely characterized this topoisomerase-DNA interaction mechanism, and hence several models for studying  it have been developed.    Proposed mechanisms  
(see \cite{v09,ldcz09} for reviews) include those 
that assume that topo II actively bends (the active bending model) or moves along the DNA (kinetic proof-reading \cite{Yan}) before
performing a strand passage
and those that assume that topo II acts preferentially at locations in the DNA that
have a specific pre-formed local conformation or ``juxtaposition'' shape, with a preference for a ``hook-like'' shape.  
To study the proposed interaction mechanisms,
various random polygon strand-passage models have been used: 
from worm-like chains to freely-jointed chains to self-avoiding lattice polygons \cite{s00,s09,fms04,fms07,hnrav07,lc08,lmzc06,lzc10,v09}.  
While the worm-like chain models are  the closest to being DNA-like,  the simpler lattice models have the advantage that the excluded volume property can be easily incorporated and that they are amenable to combinatorial and asymptotic analysis. In addition, lattice polygon models (see for example \cite{Marcone07}) can exhibit similar scaling behaviour with respect to knot localization as that observed in DNA knot experiments \cite{Ercolini07}.

One point for comparison between the models and experiments, is  the {\it knot reduction factor}, $R_K$,
introduced in \cite{lmzc06}:
\begin{equation}
R_K=\frac{P_{\bar{\phi}}^{eq}/P_\phi^{eq}}{P_{\bar{\phi}}^{st}/P_\phi^{st}}=\frac{\mbox{ratio of knots ($\bar{\phi}$) to unknots ($\phi$) at equilibrium (eq)}}{\mbox{ratio of knots ($\bar{\phi}$) to unknots ($\phi$) at steady-state (st)}},
\end{equation}
where $R_K>1$ indicates that the ratio of knots to unknots at steady state is smaller than at equilibrium and thus knotting has been reduced. For the experiments, the phrase ``thermodynamic equilibrium''
refers to the distribution of knots resulting from random cyclization of a linear duplex DNA with cohesive ends, and the phrase
``steady-state'' refers to the corresponding distribution after a topoisomerase-catalyzed reaction (with continuous ATP hydrolysis) has reached its steady state.   For a random polygon strand-passage model,  typically ``equilibrium'' refers to the distribution of knots  over all possible  random polygon conformations and ``steady-state'' refers to the knot distribution that results when transitions from one polygon conformation to another can only occur via a specified local strand-passage mechanism.  
For the simplest 2-state model where a polygon is either unknotted ($\phi$) or knotted  ($\bar{\phi}$),
the knot reduction factor reduces to \cite{lmzc06}
\begin{equation}
R_K= \frac{P_{\bar{\phi}}^{eq}t_{\bar{\phi}\to\phi}}{P_\phi^{eq}t_{\phi\to \bar{\phi}}}
\end{equation}
where $t_{a\to b}$ is the one-step transition probability for going from
state $a\in \{\phi, \bar{\phi}\}$ to state $b\in \{\phi, \bar{\phi}\}$.   Thus for a fixed equilibrium distribution,  the one-step transition probabilities for a strand-passage model determine the knot reduction factor. 
Given a random polygon model, one goal is to vary the strand-passage mechanism in order to determine factors that result in the most knot reduction; these are candidate factors for playing a role in the actual topoisomerase-DNA interaction.     In this paper, as a first step towards this, we do not calculate knot reduction factors but instead focus on the theoretical and numerical investigation of the one-step transition {\it knotting probability}
$t_{\phi\to \bar{\phi}}$ (and related quantities) for a lattice polygon model of strand passage.

In 2000  \cite{s00}, we  proposed the first lattice polygon model for studying strand passage.  Assuming a dilute solution and good solvent conditions, in \cite{s00} we consider a ring polymer  in which two segments of the polymer have already been brought close together for the purpose of performing a local strand passage. The conformations of the ring polymer are represented by self-avoiding polygons (SAPs) on the simple cubic lattice containing a specific structure $\Theta$ (located at the strand passage site - see figure \ref{graph_defn_theta} (a) in section \ref{section2}); such SAPs are referred to as $\Theta$-SAPs.  Our particular choice for the strand-passage structure $\Theta$ was motivated initially by its similarity to the Berger \etal \cite{bghw96} proposed shape for the topoisomerase-DNA complex; it should be noted, however, that the precise shape of the topoisomerase-DNA complex is still an open question.   In our model, each equal-length $\Theta$-SAP (with $\Theta$ fixed at the origin) is considered to be equally likely as a possible polymer configuration.  Consequently we do not address  how the strand passage site was identified and formed nor the effect of different solvent conditions.    In the $\Theta$-SAP model, a strand passage is performed at  $\Theta$ only if the lattice sites between its two strands are empty (see figure \ref{graph_defn_theta} (a)),  and in this  case $\Theta$ is called
$\Theta_0$ and polygons containing it are called $\Theta_0$-SAPs.  Strand passage is then performed by replacing $\Theta_0$ by the structure $\Theta_s$ as shown in Fig \ref{graph_defn_theta} (b) and the result is a lattice polygon.   

For this model, we have investigated both numerically and theoretically \cite{s00,s09} the polygon-length dependence of the knotting probability.
Specifically,    
for the $\Theta$-SAP model and a given polygon length $n$,  consider the one-step transition {\it knot probability} from the unknot to knot-type $K$:
\begin{eqnarray}
\fl t_n(\phi\to K)= \left[\frac{p_n^{\Theta_0}}{p_n(\phi)}\right]\left[\frac{p_n^{\Theta_0}(K)}{p_n^{\Theta_0}}\right]=\left[\parbox{1.2in}{ prob.  $\Theta_0$ occurs at strand passage site}\right]\left[\parbox{1.5in}{prob. strand passage at $\Theta_0$ results in knot $K$}\right],
\end{eqnarray}
where $p_n(\phi)$ is the number of $n$-edge unknotted SAPs rooted at the origin, $p_n^{\Theta_0}$ is the number of these that contain $\Theta_0$, and $p_n^{\Theta_0}( K)$ is the
number of the latter that yield a knot-type $K$ SAP after a single strand passage is performed at $\Theta_0$.   $t_n(\phi\to \bar{\phi})$ (needed for the denominator of $R_K$) is then given by
$1-t_n(\phi\to\phi)$.   Alternatively, for the restricted equilibrium in which only $\Theta_0$-SAPs 
are considered, the {\it $\Theta_0$-restricted knot probability}, $\rho^{\Theta_0}_n(K)$, is given by the second ratio  above, namely:
\begin{equation}
\rho^{\Theta_0}_n(K)=\frac{p_n^{\Theta_0}(K)}{p_n^{\Theta_0}}.
\label{rhodef}
\end{equation}
In either case, $K$ must either be the unknot or an unknotting number one knot,
and we denote the set of all such $K$ by ${\cal K}$.
In \cite{s09}, combinatorial bounds are proved which relate the polygon counts just defined. 
These bounds yield that the number of $n$-edge unknotted $\Theta$-SAPs and $\Theta_0$-SAPs each grow exponentially (with $n$) at the same rate as the total number of $n$-edge unknotted self-avoiding polygons. Thus, for example,
$\lim_{n\to\infty} n^{-1}\log p_n(\phi)=\lim_{n\to\infty} n^{-1}\log p_n^{\Theta_0}$.  Furthermore, it is proved that the same holds for each subset of $n$-edge unknotted $\Theta_0$-SAPs that
yields a specific after-strand-passage knot-type.  Thus, for example, the {\it $\Theta_0$-restricted knotting probability}, $\rho_n^{\Theta_0}(\bar{\phi})=1-\rho_n^{\Theta_0}(\phi)$,  does not grow exponentially with $n$.  Based on a heuristic argument, it is conjectured that the leading asymptotic form (as $n$ goes to infinity) of $p_n(\phi)$,  
$p_n^{\Theta_0}$ and 
${p_n^{\Theta_0}(K)}$ are all the same, up to a positive constant, and  hence 
$0<\lim_{n\to\infty} \rho^{\Theta_0}_n(\bar{\phi}) < 1$ and $0<\lim_{n\to\infty} t_n(\phi\to \bar{\phi}) < 1$.

In \cite{s00},  a composite Markov chain (CMC) Monte Carlo (also known as multiple Markov chain) algorithm was developed for studying $\Theta$-SAPs with any given fixed knot-type. This algorithm, called the CMC $\Theta$-BFACF algorithm, is based on the BFACF algorithm \cite{bf81,cc83,ccf83}.  Also in \cite{s00},  an ergodicity proof was given for  the $\Theta$-BFACF algorithm which was a non-trivial extension of the Janse van Rensburg and Whittington \cite{jw91} ergodicity proof  for the BFACF algorithm.   
Most recently, in \cite{s09}, improved statistical methods are developed for estimating the length-dependence of the knot  probabilities from CMC Monte Carlo data  and then Monte Carlo data is used to investigate the  conjectures discussed above in relation to equation (\ref{rhodef}).   The approaches developed to study the
$\Theta$-SAP model in \cite{s09} are expected to be broadly applicable to any self-avoiding polygon strand passage model.  
In this paper, we  review the theoretical results and numerical/statistical methods from \cite{s09} and  present new results based on additional Monte Carlo data beyond that used in \cite{s09}.

Since 2000, there have been a number of different research groups investigating the one-step transition knot probabilities (e.g. \cite{fms04,fms07,hnrav07,lc08,lmzc06,lzc10,v09}) for a variety
of (on- and off-lattice) strand-passage models.  The main advantage of the $\Theta$-SAP model over the newer models is that it has been possible to prove results about it. In contrast, little if anything has been proved about the newer models and each model has aspects (e.g. off-lattice polygons or virtual strand passages) which make mathematical rigour a challenge.
The newer strand passage models, \cite{fms04,fms07,hnrav07,lc08,lmzc06,lzc10,v09}, and their studies have, however, raised a number of important questions.  The most important of these with respect to modelling DNA is:  How do the  knot probabilities depend on the local juxtaposition geometry around the strand-passage site?  For example,
from the  strand-passage model studies of \cite{lmzc06,lzc10}, it is observed that  the ``tightness'' or ``compactness'' of the local juxtaposition geometry affects the knot reduction factor and the knotting probabilities.  
Specifically, in 2006, Liu \etal \cite{lmzc06} investigated
knotting probabilities after a local ``virtual''  strand passage in SAPs in $\mathbb{Z}^{3}$; the strand passage is termed virtual since the after-strand-passage polygon need not be a SAP in $\mathbb{Z}^{3}$. 
They investigated knotting (and unknotting) probabilities as a function of the local juxtaposition geometry of the two polygon segments involved in the virtual strand passage,  and highlighted their results for three classes of juxtapositions (from most  to least compact): ``hooked'', ``half-hooked'', and ``free'' (cf.
\cite[table 1]{lmzc06}).
They found
that strand passages about a hooked juxtaposition (when compared
to the other two types)
had the lowest knotting probability (essentially zero) and those about\ a free\ juxtaposition
had the highest. \ In \cite{lzc10}, they obtained similar conclusions for an
off-lattice model; specifically, from \cite[table I]{lzc10}, the probabilities of knotting for hooked, half-hooked, and free juxtapositions are reported to be, respectively, 0.0028, 0.0077, and 0.1014.

The structure $\Theta$ resembles the half-hooked juxtaposition of \cite{lmzc06}, but it is not the same.  Furthermore, for the $\Theta$-SAP model, we only consider a strand passage that
yields a SAP in $\mathbb{Z}^{3}$.
Thus it is not possible to directly compare the Liu \etal results to any from the $\Theta$-SAP model.  However, it is
possible to investigate how the knotting probabilities for $\Theta_0$-SAPs
depend on the local geometry immediately adjacent to the fixed structure
$\Theta_0$.  Based on the Liu \etal results, it is expected that
the knotting probabilities do depend on this local geometry. For the $\Theta$-SAP model, the {\it geometry-dependent one-step transition knot probability} for $K\in {\cal K}$ is given by:
\begin{equation}
t_n^{G}(\phi\to K)= \left[\frac{p_n^{G}}{p_n(\phi)}\right]\left[\frac{p_n^{G}(K)}{p_n^{G}}\right],
\end{equation}
where $p_n^{G}$ is the number of $n$-edge unknotted $\Theta_0$-SAPs that have a specified local juxtaposition geometry $G$ and $p_n^{G}(K)$ is the number of these that yield a knot-type-$K$ SAP after a single strand passage is performed at $\Theta_0$.    Alternatively, for the restricted equilibrium in which only $\Theta_0$-SAPs with local geometry $G$
are considered, the {\it $G$-restricted knot probability}, $\rho_n^{G}(K)$, is given by the second ratio  above, namely:
\begin{equation}
\rho_n^{G}(K)=\frac{p_n^{G}(K)}{p_n^{G}}.
\end{equation}
For our numerical investigation, we focus on $\rho_n^G(\bar{\phi})=1-\rho_n^{G}(\phi)$ and explore its dependence on $G$.
Because the hooked, half-hooked, and free local geometry classifications do not translate to the $\Theta$-SAP model, we need alternate schemes to classify  the local geometry.
In this paper, we propose two new compactness classification schemes and show that the corresponding knotting probabilities decrease with compactness in each case.  In both schemes, we find that the sign of the crossing at the strand passage site plays a significant role.    This latter observation is consistent with
experimental work on DNA which indicates that some topoisomerases  exhibit a chirality bias \cite{neuman,ShawWang97,Roca,LynnIMA}.

The purpose of this paper is thus two-fold. 
Our first goal is to summarize the theoretical results and conjectures and the numerical methods from
\cite{s09} regarding the knot probabilities.   Using these methods, the conjectures are then tested, based on new data beyond that which was available in \cite{s09}.
Our second goal is to investigate, for the first time, the effect of the local juxtaposition geometry on the  $\Theta$-SAP model  knotting probabilities.  To do this, we first extend the combinatorial arguments developed in \cite{s09}
to obtain analogous results and conjectures about the asymptotic properties of $\rho_n^{G}(K)$.   We then use the numerical methods of \cite{s09} to investigate numerically the dependence of $\rho_n^G(\bar{\phi})$ on $n$ and $G$.  We establish first that it is highly dependent on the crossing-sign  at the strand passage site.  
 In order to determine which factors are most influential on the 
knotting probability, we investigate two geometric properties of the juxtapositions - an ``opening'' angle and a
``compactness'' measure.
We find a  trend for ``compactness'' versus knotting probability which is consistent with results obtained for other strand-passage models \cite{lmzc06,lzc10}.
We also find an ``opening'' angle versus knotting probability trend which is noteworthy in light of
recent experimental results \cite{neuman}.  In particular, our ``opening'' angle is defined to be consistent with the angle defined in \cite{neuman}.  They find that topo IV (a type II topoisomerase) binds preferentially when this angle is slightly acute; we find that the more acute the angle,  the lower the knotting probability.

In summary, in the next section of the paper (section \ref{section2}), the  $\Theta$-SAP model for local strand passage in  unknotted ring polymers is defined.  We then investigate, both theoretically and numerically, the distribution of knots obtained after performing a strand passage in an unknotted $n$-edge $\Theta_0$-SAP.  
Given $K\in{\cal{K}}$, we heuristically argue (in section \ref{asymsec}) that its probability of resulting from a strand passage at $\Theta_0$ depends on $n$ and approaches (as $n\to\infty$) a limit lying strictly between 0 and 1.  We prove (in section \ref{asymsec}) that the rate of approach to the limit is less than exponential. In  order to investigate the knot distribution further as a function of polygon length, the CMC $\Theta$-BFACF algorithm is used. The ergodicity classes for this algorithm are discussed (see section \ref{thetabfacfsection}). Then, the maximum likelihood estimation approach for analyzing CMC data from \cite{s09} is reviewed (see section \ref{mlesec} and appendix A) and used to provide evidence supporting our heuristic arguments (see section \ref{mlesec}). Then our best estimates of the $\Theta_0$-restricted knot probabilities are presented (see section \ref{PROBABILITY_SECTION}).    Finally, we present evidence that
the probability of going from an unknot to a knot for $\Theta_0$-SAPs does depend on the local structure around the strand-passage site and especially on the crossing-sign at the strand-passage site (see section \ref{juxtasec}).

\section{The unknotted \boldmath{$\Theta$}-SAP model}
\label{section2}

\begin{figure}[htb]
\centering\psfrag{A}{\small{$A$}}
\psfrag{B}{\vspace{0.2in}\hspace{-0.05in}\small{$B$}} \psfrag{C}{\small{$C$}}
\psfrag{D}{\hspace{-0.1in}\small{$D$}}
\psfrag{E}{\hspace{-0.02in}\small{$E$}} \psfrag{F}{\hspace{-0.02in}\small
{$F$}} \psfrag{G}{\hspace{-0.02in}\small{$G$}} \psfrag{H}{\small{$H$}}
\mbox{\subfigure[{}]{
\epsfig{file=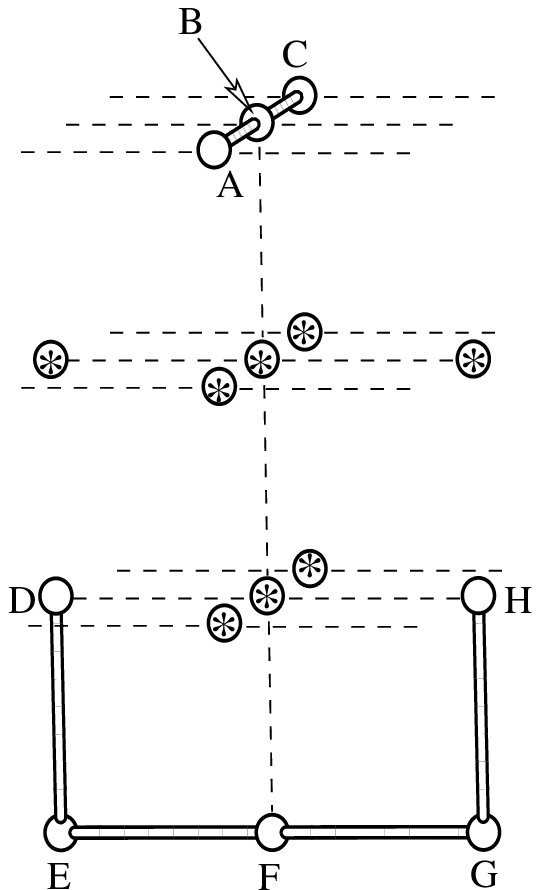,scale=0.50,angle=0}
}
\subfigure[{}]{
\epsfig{file=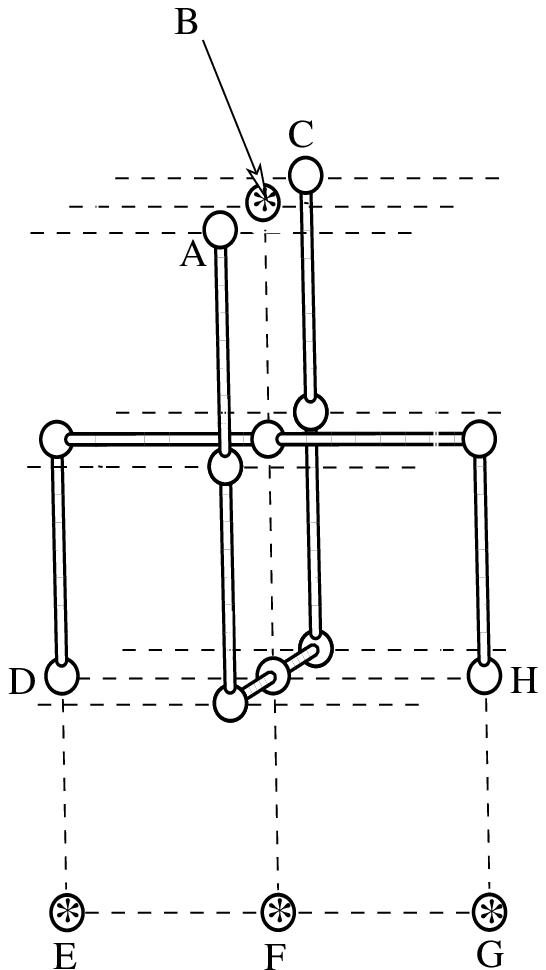,scale=0.50,angle=0}}
\includegraphics[scale=0.7
]
{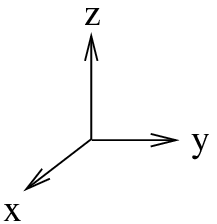}
}
\psfrag{A}{\hspace{-0.05in}\small{$A$}}
\psfrag{B}{\small{$B$}} \psfrag{C}{\small{$C$}}
\psfrag{D}{\hspace{-0.1in}\small{$D$}}
\psfrag{E}{\hspace{-0.02in}\small{$E$}} \psfrag{F}{\hspace{-0.02in}\small
{$F$}} \psfrag{G}{\hspace{-0.02in}\small{$G$}} \psfrag{H}{\small{$H$}}
\mbox{\subfigure[{}]{\includegraphics[scale=0.5
]{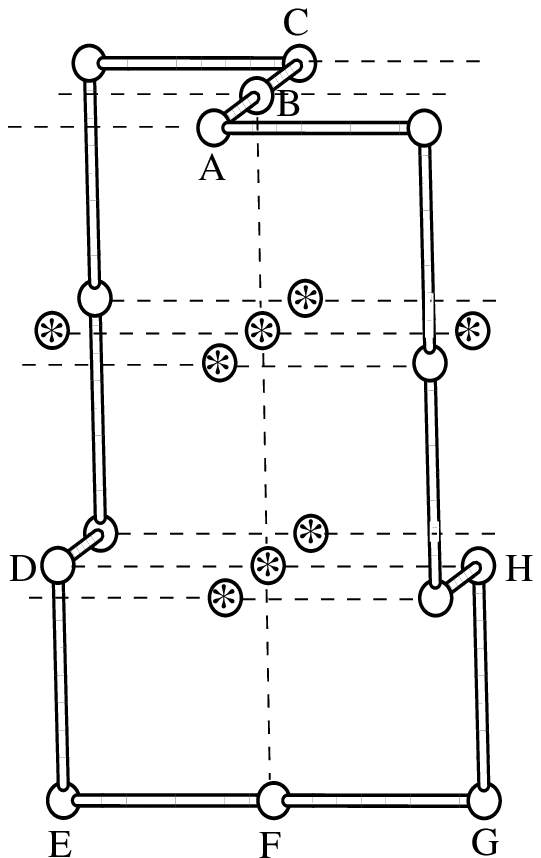}}
\subfigure[{}]{
\includegraphics[scale=0.5
]{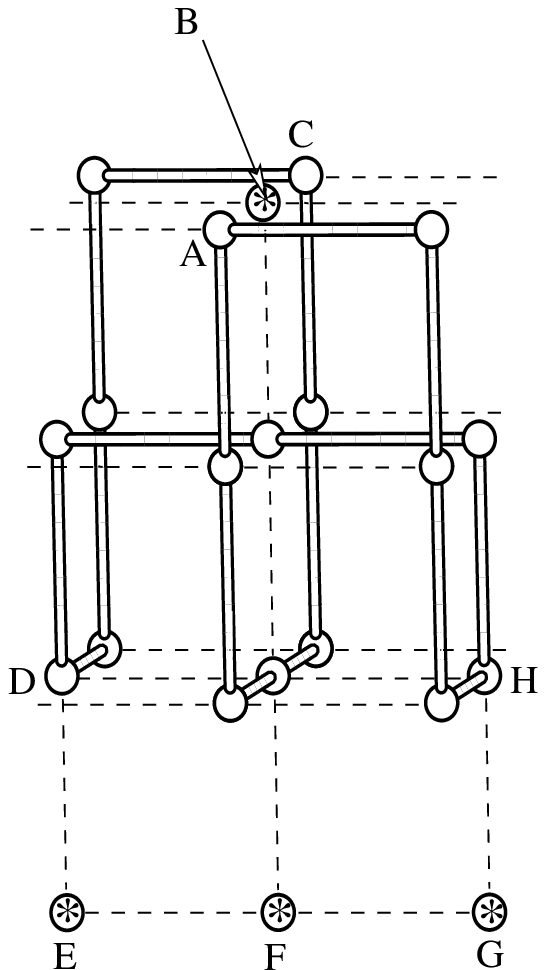}
}
\psfrag{A}{\hspace{-0.05in}\small{$A$}}
\psfrag{B}{\small{$B$}} \psfrag{C}{\small{$C$}}
\psfrag{D}{\hspace{-0.1in}\small{$D$}}
\psfrag{E}{\hspace{-0.02in}\small{$E$}} \psfrag{F}{\hspace{-0.02in}\small
{$F$}} \psfrag{G}{\hspace{-0.02in}\small{$G$}} \psfrag{H}{\small{$H$}}
\subfigure[{}]{\includegraphics[scale=0.5
]{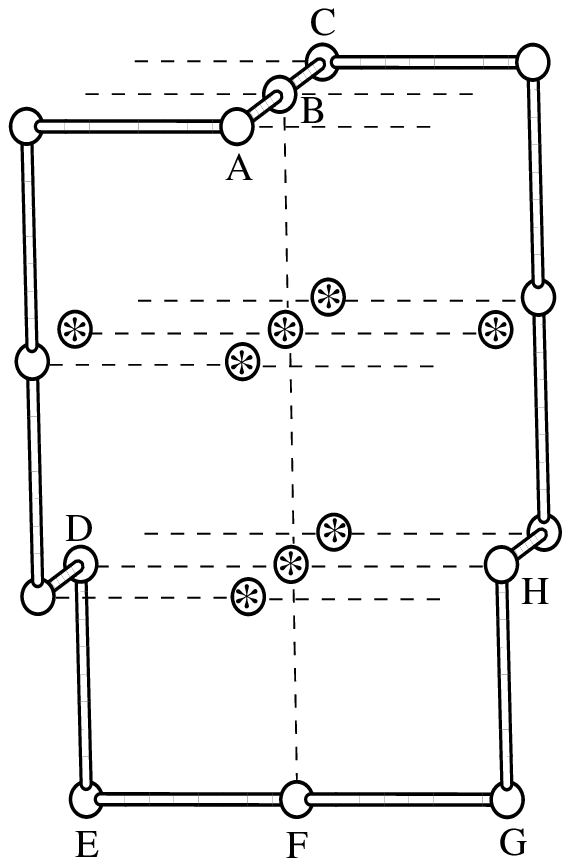}
}}
\caption{ (a) The fixed strand passage structure $\Theta$:  open and empty
circles represent its vertices  and open bonds represent its edges.  \ Dashed lines and circles containing asterisks represent,
respectively, lattice edges and vertices that $\Theta$ does not occupy but which may be occupied in a $\Theta$-SAP. \  In the case that the circles containing asterisks are not allowed to be occupied, the strand passage structure is called $\Theta_0$.
 \  $A=(1,0,0)$; $B=(0,0,0)$; 
$C=(-1,0,0)$;  $D=(0,-1,-2)$; $E=(0,-1,-3)$; 
$F=(0,0,-3)$; $G=(0,1,-3)$; and $H=(0,1,-2)$.
(b) The after-strand-passage structure $\Theta_{s}$:  open and
empty circles represent its vertices  and open bonds represent
its edges. \ The circles containing asterisks are vertices in
$\Theta$ not occupied by $\Theta_{s}.$ 
(c) An unknotted 14-edge $\Theta_0$-SAP $\omega$ and (d) the corresponding 18-edge after-strand-passage polygon $\omega_s$. (e) shows the $\Theta^+_0$-SAP $\widetilde{\omega}$ obtained via the mirror operation $\widetilde{~~}$ from the $\Theta^-_0$-SAP $\omega$ of (c). 
 }
\label{graph_defn_theta}
\end{figure}

An $n$-edge
self-avoiding polygon (SAP or polygon, for short)  is an $n$-edge connected subgraph on the simple cubic lattice
$\mathbb{Z}^3$ with each vertex having degree two.  
For SAPs, the number of edges, $n$, must be greater than
$3$ and even and this will be assumed henceforth. 
In our model, we assume that two strands of the polymer have already been  \textquotedblleft
pinched\textquotedblright\ together  for the purpose of implementing a strand passage.
\label{sec_str_psg_model}To model the
pinched portion of the ring polymer, the SAPs used are
required to contain the pattern $\Theta$  (as illustrated in figure~\ref{graph_defn_theta} (a))
fixed at $(0,0,0)\in\mathbb{Z}^{3}$.
The  lattice space between the two strands of $\Theta$ is needed to ensure that all $\Theta$-SAPs can be generated via
a sequence of BFACF moves (this is explained further in section \ref{thetabfacfsection}).

To perform a strand passage at $\Theta$ in a $\Theta$-SAP,  $\Theta$ is replaced by the fixed \textit{after-strand-passage structure} $\Theta_{s}$ as illustrated in figure~\ref{graph_defn_theta} (b). 
The vertices  in\ figure \ref{graph_defn_theta} (a) that are represented by circles containing
asterisks must not be end points of any edge of the initial polygon in order for this strand passage to yield a lattice polygon; hence strand passage is only performed at $\Theta_0$.  
For a $\Theta_0$-SAP $\omega$, the polygon, $\omega_s$, obtained by replacing $\Theta$ with $\Theta_{s}$  is referred to as the resulting
\textit{after-strand-passage polygon}.
(Figure \ref{graph_defn_theta} (c) is an illustration of a 14-edge $\Theta_0$-SAP  and
Figure \ref{graph_defn_theta} (d) is an illustration of the after-strand-passage polygon obtained from it.)
When $\omega_s$ has knot-type $K$, then $\omega$ is called a $\Theta_0(K)$-SAP.

Note that $\omega_s$ has four more edges than $\omega$; the extra edges ensure that $\omega_s$ is a lattice polygon.  Thus, strictly speaking, the transition knot probabilities calculated using this model will not correspond to those for a realizable steady state of fixed polygon-length lattice SAPs.  However, we argue that this is no worse than for the lattice strand-passage models of \cite{lmzc06,hnrav07} which do not  yield an after-strand-passage lattice
polygon. Furthermore, just as for these other models, we expect the $\Theta$-SAP model will predict 
qualitative trends that are broadly applicable.  In addition, we mitigate this problem somewhat by considering averages over polygons of varying lengths when calculating knot probabilities (see section \ref{PROBABILITY_SECTION}). 

For the remainder of the paper, our focus is on unknotted $\Theta$-SAPs only and, unless stated otherwise, the term $\Theta$- or $\Theta_0$-SAP refers to only those that are unknotted.

\begin{figure}[hhh]
\centering\psfrag{A}{$A$} \psfrag{B}{$B$} \psfrag{C}{$C$}
\psfrag{D}{$D$} \psfrag{E}{$E$} \psfrag{F}{$F$} \psfrag{G}{$G$}
\psfrag{H}{$H$} \psfrag{I}{$I$} \psfrag{J}{$J$}
\includegraphics[scale=0.5]{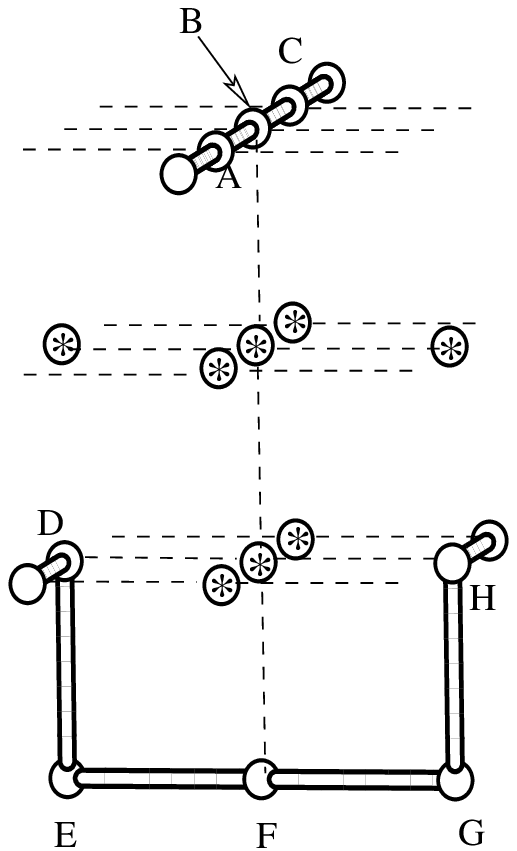}\includegraphics[scale=0.5]{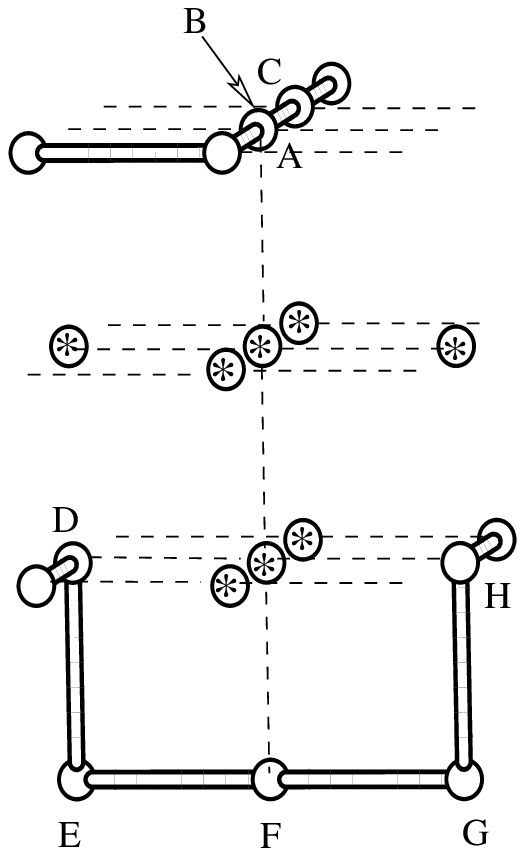}\includegraphics[scale=0.5]{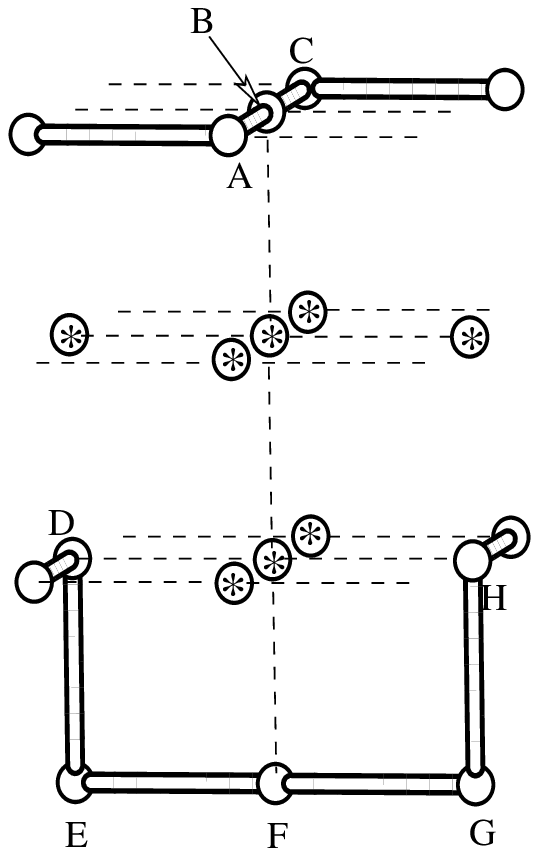}
\caption{Illustrations of example juxtapositions: For the sake of simplicity we refer to these juxtapositions, from left to right, as $S$ (straight top), $L$ (L-top), and $Z$ (Z-top).
}
\label{free_juxtaposition copy(2)-1-1}
\end{figure}

The polygon conformation around $\Theta$ will be used to investigate juxtaposition-geometry-effects on knotting.  To do this, 
we define a $\Theta_0$-SAP $\omega$'s juxtaposition,  $J$, by the  vertices $(v_1, v_2, v_3, v_4)$  of $\omega$ that are not in $\Theta$ but are, respectively, immediately adjacent to the vertices $A$, $C$, $D$ and $H$ of $\Theta$.  In this case, $\omega$ is called a $J$-SAP and, when $\omega_s$ has knot-type $K$, a $J(K)$-SAP.  There are 144 juxtapositions $J$ that can occur in a $\Theta_0$-SAP and we denote the set of these juxtapositions by ${\cal{J}}$. See figure \ref{free_juxtaposition copy(2)-1-1} for examples of $J\in{\cal J}$. Since we will refer to these examples further, we name them according to the shape of the top segment of the juxtaposition. That is, we name them respectively (from left to right in figure \ref{free_juxtaposition copy(2)-1-1})
as $S$ (for straight top), $L$ (for L-shaped top) and $Z$ (for Z-shaped top). 

\begin{figure}[hhh]
\centering
\mbox{\subfigure[{}]{
\includegraphics[scale=0.5]{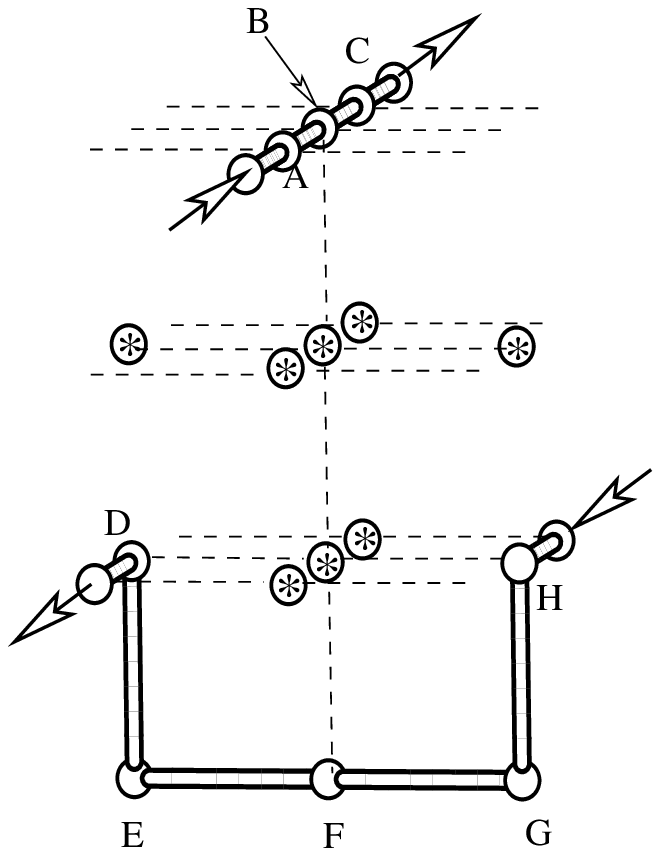}}\subfigure[{}]{\includegraphics[scale=0.5]{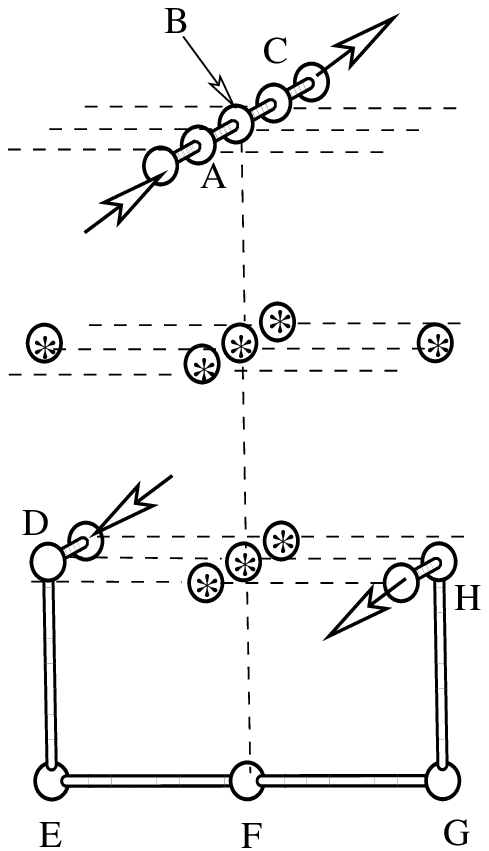}}\subfigure[{}]{\includegraphics[scale=0.5]{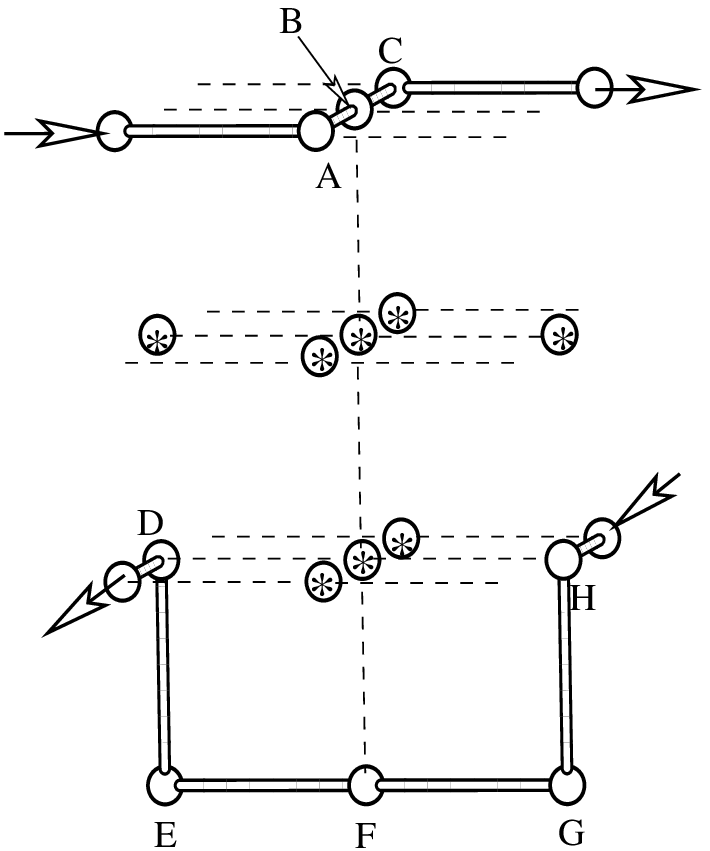}}\subfigure[{}]{\includegraphics[scale=0.5]{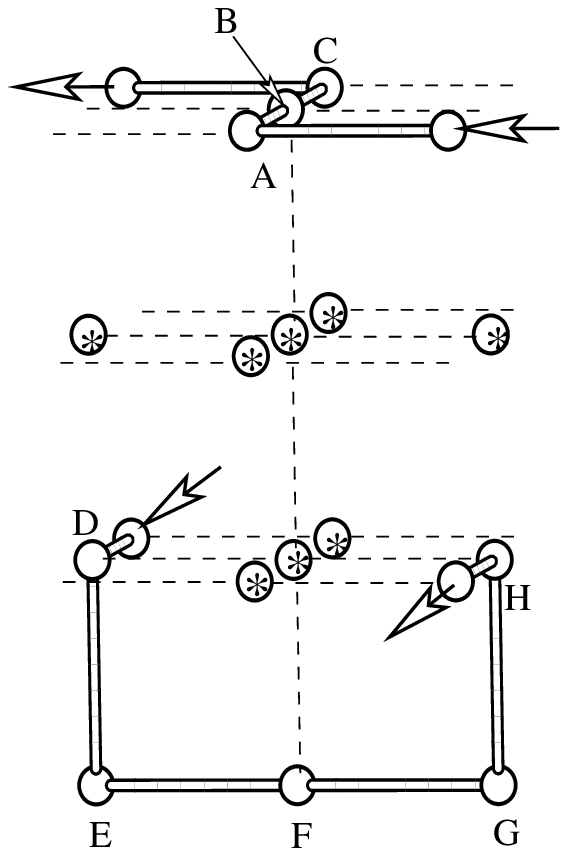}}
}
\mbox{\subfigure[{}]{
\includegraphics[scale=0.5]{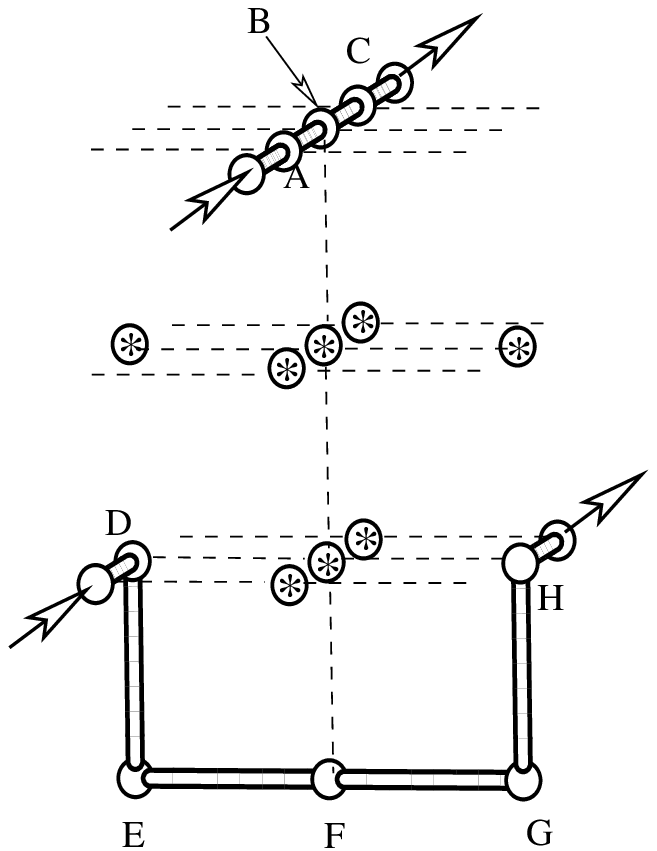}}\subfigure[{}]{\includegraphics[scale=0.5]{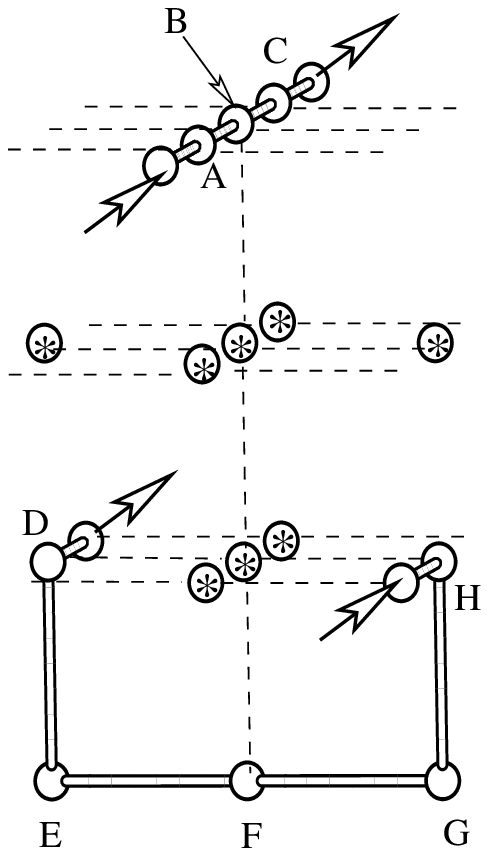}}\subfigure[{}]{\includegraphics[scale=0.5]{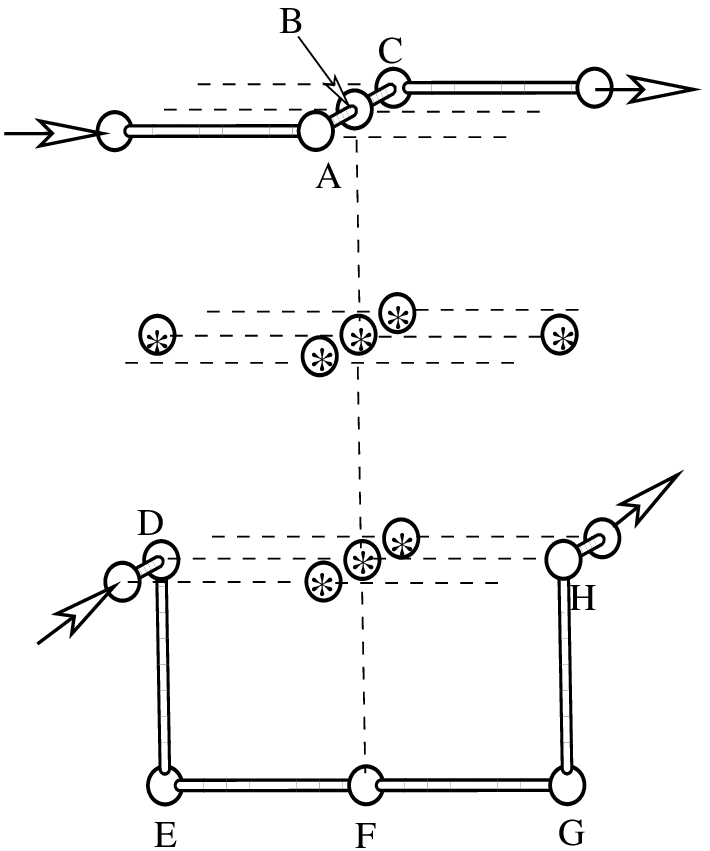}}\subfigure[{}]{\includegraphics[scale=0.5]{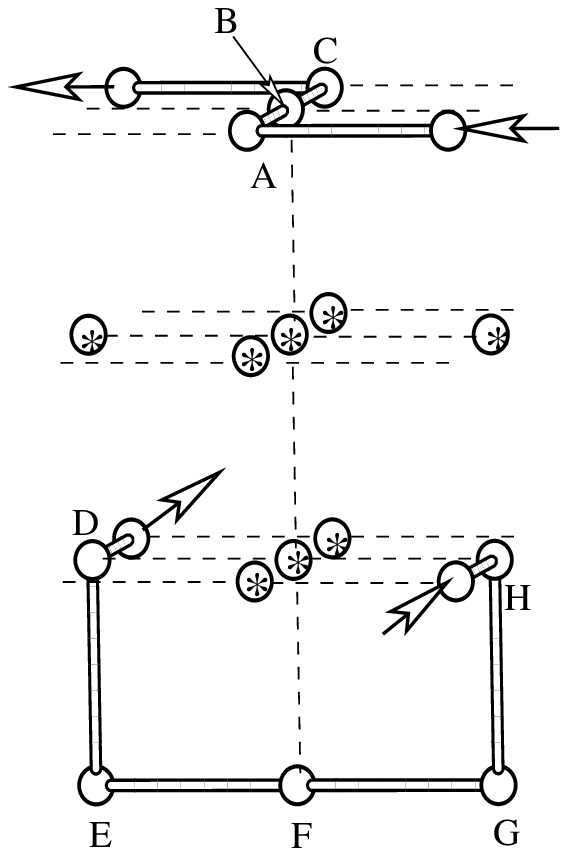}}}
\caption{ Illustrations of juxtapositions (a) $S^+$, (b) $\widetilde{S^+}$ ($S^+$-mirror), (c) $Z^+$, and (d) $\widetilde{Z^+}$ ($Z^+$-mirror), (e)  $S^-$, (f) $\widetilde{S^-}$, (g) $Z^-$, and (h) $\widetilde{Z^-}$ respectively.}
\label{free_juxtaposition copy(2)-1-1b}
\end{figure}

Depending on how the endpoints of $\Theta$ are paired and joined to form the
polygon $\omega$,  the projection of $\Theta$ into the $z=0$ plane results in either a positively (+) or negatively (-) signed crossing (according to a right-hand-rule).   In the former case,  outside $\Theta$, $C$ is always directly connected to $H$ in $\omega$, while in the latter case $C$ is always directly connected to $D$.   For the (+) case, $\Theta$ is labelled $\Theta^+$, $\omega$'s
juxtaposition is labelled $J^+$ and $\omega$ is called a $\Theta^+$-SAP and a $J^+$-SAP;  in the (-) case, $\Theta$ is $\Theta^-$, the juxtaposition is $J^-$ and $\omega$ is a $\Theta^-$-
and a $J^-$-SAP.     Thus each $J\in{\cal{J}}$  has a $(+)$ and $(-)$ version.  For each $\sigma\in\{+,-\}$, we use $J^{\sigma}(K)$-SAP to refer to any $J^{\sigma}$-SAP whose after-strand-passage polygon has knot-type $K\in{\cal{K}}$.
Figure \ref{free_juxtaposition copy(2)-1-1b} displays examples of signed
juxtapositions; the arrows indicate how the end points are joined in any polygon containing the juxtaposition (by convention, we always orient the top strand of $\Theta$ from $A$ to $C$).  

Depending on the extent to which the local geometry, $G$, at the strand passage site is specified,
we can define associated polygon counts.  That is, for each $G\in\{\Theta,\Theta^+,\Theta^-,\Theta_0,\Theta_0^+,\Theta_0^-,J,J^+,J^-:J\in{\cal{J}}\}$, $p_n^G$ and $p_n^G(K)$ are defined respectively as the number of $n$-edge $G$-SAPs and $G(K)$-SAPs.  \ Define also $p_{n}^{G}(\bar{\phi}):=p_{n}^{G}- p_{n}^{G}({\phi})$
and $p_{n}^{G}(\bar{\phi}):=p_{n}^{G}- p_{n}^{G}({\phi})$.

In addition, consider the reflection or ``mirror'' operation ``$\widetilde{~~}$'' which takes unoriented $(x,y,z)\in\mathbb{Z}^3$ to $(-x,y,z)\in\mathbb{Z}^3$.  
Figures~\ref{graph_defn_theta} (c) and (e)  illustrate  two 14-edge $\Theta_0$-SAPs that are related via this reflection.   In fact $\widetilde{~~}$ provides a one-to-one mapping between the sets of $\Theta^+$- and $\Theta^-$-SAPs and, for example, 
each juxtaposition $J^-$ corresponds to a unique $(+)$ juxtaposition $J^-$-mirror given by $\widetilde{J^-}$.  
(Figure \ref{free_juxtaposition copy(2)-1-1b} displays examples of  juxtaposition pairs related by  $\widetilde{~~}$.   Note that there are only 4 juxtapositions
for which $J^+=\widetilde{J^-}$. )
Furthermore,  given any $K\in{\cal K}$, for any $J^-(K)$-SAP $\omega$, for example, with after-strand-passage polygon $\omega_s$, the after-strand-passage polygon $\widetilde{\omega_s}$  of  the $\Theta^+_0$-SAP  $\widetilde{\omega}$  (a $\widetilde{J^-}$-SAP) has the same knot-type $K$ as that of $\omega_s$,  except with the opposite chirality in the case that $K$ is chiral.   
Thus (ignoring chirality) we have that
\begin{equation}
p_n^{\Theta_0^+}(K)=p_n^{\Theta_0^-}(K)=p_n^{\Theta_0}(K)/2;
\label{connectionclassrelation}
\end{equation}
and for the signed-juxtaposition polygon counts,
\begin{eqnarray}
p_n^{J}(K)&=&p_n^{J^+}(K)+p_n^{J^-}(K)=
p_n^{J^+}(K)+p_n^{\widetilde{J^-}}(K)
=p_n^{J^-}(K)+p_n^{\widetilde{J^+}}(K).
\label{juxtconn}
\end{eqnarray}
Thus the juxtaposition-specific after-strand-passage knot probabilities
can be determined by focussing on polygons of only one crossing-sign type.  We will rely on this fact for our simulation results.

Given that there are 144 different
juxtapositions,  it is useful to group the juxtapositions further.  Motivated by the fact that the
``hooked" juxtaposition of \cite{lmzc06} is more compact than their ``free'' juxtapositions, we
use two classification schemes to measure the ``compactness" or ``tightness" of a juxtaposition.  The first scheme proposed here is to use the size, denoted $l(G)$, of the smallest
$\Theta_0$-SAP that can contain a specified juxtaposition geometry  $G$
to measure its compactness. For example, for $G$ given by a single juxtaposition $J$  (or $J^+$ or $J^-$), $G$ 
is said to be a compactness  size-$m$  (or $m^+$ or $m^-$) juxtaposition if $l(G)=m$ and any  $\Theta_0$-SAP that contains it is referred to as an $m$- (or $m^+$- or $m^-$-) SAP.
Note that $m\in {\cal{M}}=\{14,16,18,20,22\}.$  Also, any  $G$-SAP,  $G\in\{m,m^+,m^-\}$, whose
after-strand-passage polygon has knot-type $K\in{\cal{K}}$ is called a $G(K)$-SAP.

For all but 36 of the unsigned juxtapositions $J\in{\cal J}$,  $l(J)$
is equal to only one of $l(J^+)$ or $l(J^-)$.   For example, a smallest (size-14) $\Theta_0$-SAP that contains juxtaposition $Z$ (as illustrated in figure~\ref{free_juxtaposition copy(2)-1-1}) must contain $Z^+$, and any  $\Theta_0$-SAP containing
$Z^-$ must have more than 14 edges (in fact at least 22 edges).  Consequently juxtaposition $Z^-$ is not as ``compact'' (by our definition)
 as $Z^+$.  Hence the crossing sign can play a role in how small (compact) a polygon
can be that contains a particular juxtaposition.

\begin{figure}[hhh]
\centering\psfrag{LT}{\hspace{-0.2in}\small{$L_{12}$}}\psfrag{LB}{\hspace{-.2in}\small{$L_{43}$}}
\mbox{\subfigure[{}]{
\includegraphics[scale=0.5]{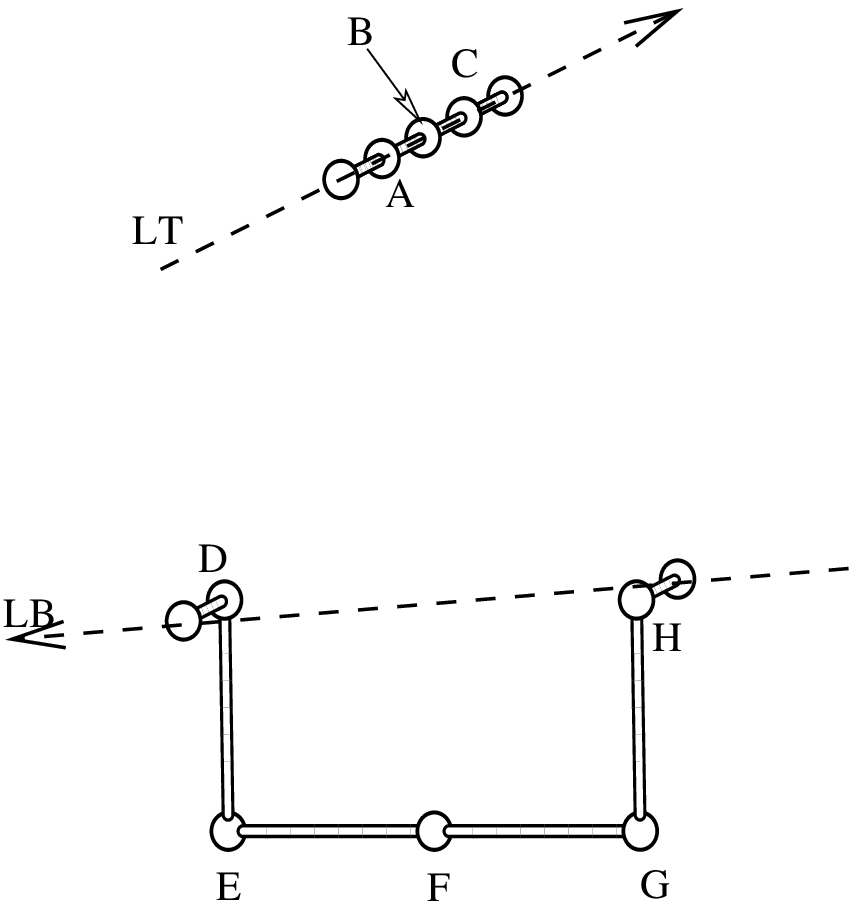}}\hspace{0.5in}\psfrag{LT}{\small $l_{12}$}\psfrag{LB}{\small $l_{43}$}\psfrag{v1}{\tiny{$P_1$}}\psfrag{v2}{\tiny{$P_2$}}\psfrag{v3}{\hspace{-0.1in}\tiny{$P_3$}}\psfrag{v4}{\hspace{-0.1in}\tiny{$P_4$}}\psfrag{A}{\tiny{$P_A$}}\psfrag{B}{\tiny{$P_B=I$}}\psfrag{C}{\tiny{$P_C$}}\psfrag{D}{\hspace{-.1in}\tiny{$P_D$}}\psfrag{H}{\tiny{$P_H$}}\psfrag{al}{$\alpha$}\subfigure[{}]{\includegraphics[scale=0.5]{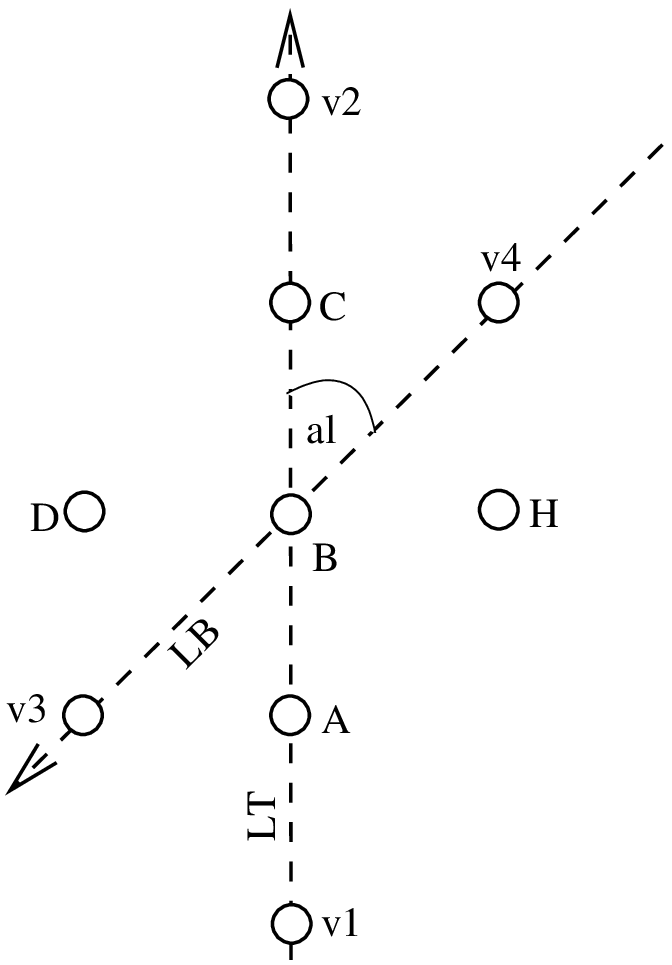}
}}
\caption{(a) Juxtaposition $S^+$; the oriented dashed lines $L_{12}$ and $L_{43}$ project, respectively, to $l_{12}$ and $l_{43}$ in the $z=0$ plane as shown in (b).  (b) The opening angle $\alpha$ associated with juxtaposition $S^+$.}
\label{angle_def}
\end{figure}

\begin{figure}[hhh]
\centering\psfrag{259}{\tiny{$\widetilde{S^-}$}} \psfrag{252}{\hspace{-0.05in}\tiny{$S^{+}$}} \psfrag{511}{\hspace{-0.05in}\tiny{$\widetilde{Z^-}$}}
\psfrag{756}{\hspace{-0.05in}\tiny{$Z^{+}$}} \psfrag{class}{\hspace{-.1in}\tiny{$m$}} \psfrag{angle}{\small{$\alpha$}}
\includegraphics[scale=1]{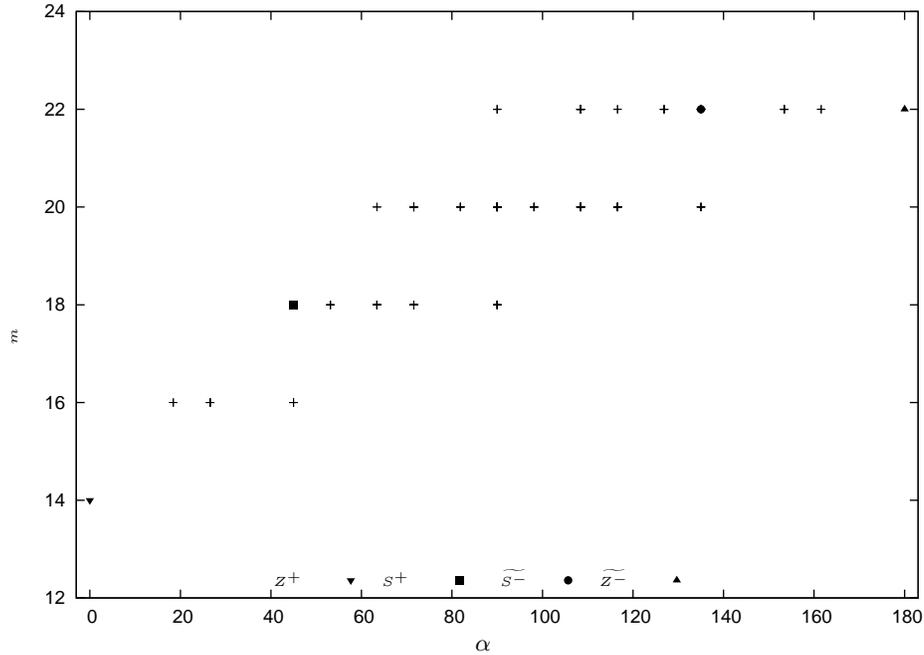}
\caption{Correlation between the compactness-size $m^+$ and opening angle $\alpha^+$ for the 144 (+) juxtapositions.}
\label{ADD}
\end{figure}

As another measure of compactness, 
we also define an {\it opening angle} associated with each signed juxtaposition.
The angle is defined to be consistent with \cite[Fig. 1 A]{neuman} in the positive supercoil case.  
Given a juxtaposition $J^{+}$ (determined by $v_1, v_2, v_3, v_4$), consider its projection onto the $z=0$ plane. \ Under this projection,
denote the image of the points $A,B,C,D,H, v_{1},v_{2},v_{3},$ and $v_{4},$
respectively  to be $P_{A},P_B,P_{C},P_{D},P_{H},P_1,P_2,P_3$ and $P_{4}.$ \ Now, in the $z=0$ plane, consider
the  directed line ($l_{12}$) from $P_{1}$ to $P_{2}$ and the  directed line
($l_{43}$) from $P_{4}$ to $P_{3}.$ (Note that the directions on these lines are assigned to be consistent with a positively signed crossing.)  \ These two lines have at least one point of intersection, choose one such point and label it $I$. Now form two rays, $r_{I,2}$ and
$r_{I,4}$ where $r_{I,2}$ starts at $I$ and follows $l_{12}$ to $P_2$ and $r_{I,4}$ starts at $I$ and follows $-l_{43}$ to $P_4$.
Rays $r_{I,2}$ and $r_{I,4}$ together define the {\it  opening angle} for juxtaposition $J^+$
with the initial
leg of the angle  $r_{I,4}$ and the terminal leg 
$r_{I,2}$.  (See for example figure \ref{angle_def} with $J^{+}=S^{+}$.) 
Outside $\Theta$, $v_2$ is joined next to $v_4$ in any polygon containing $J^+$;  the opening angle is thus one measure of how ``far'' these two vertices are apart in such a polygon.  Roughly speaking, a larger opening angle yields more space between $v_2$ and $v_4$ and we view the juxtaposition as being more ``open''.    In fact, as shown in figure \ref{ADD}, there is a correlation between the compactness-size of  $J^+$ and its opening angle.
The opening angle for a juxtaposition $J^{-}$ is obtained by subtracting the opening angle of $J^{+}$ from 180$^{\circ}$;  this ensures that  $J^+$ and its mirror, $\widetilde{J+}$,   have the same opening angle.

Using this technique, the possible opening angles for the $\Theta_0$-SAP
signed juxtapositions  are in the set ${\cal{A}}=\{$ $0^{\circ}$, $18.43^{\circ}$, $26.57^{\circ}$, $45^{\circ}$, $53.13^{\circ}$, $63.43^{\circ}$, $71.57^{\circ}$, $81.87^{\circ}$, $90^{\circ}$,  $98.13^{\circ}$, $108.43^{\circ}$, $116.57^{\circ}$, $126.87^{\circ}$, $135^{\circ}$, $153.43^{\circ}$, $161.57^{\circ}$, and $180^{\circ}\}$.
For each $\sigma\in\{+,-\}$, any $\Theta_0^\sigma$-SAP with opening angle $\alpha$ is called an $\alpha^\sigma$-SAP, and if its
after-strand-passage polygon has knot-type $K$, it is an $\alpha^\sigma(K)$-SAP.

To investigate the probability of knotting
as a function of polygon length and juxtaposition compactness or juxtaposition opening angle, we use the associated polygon counts.    That is, for $G\in\{m,m^+,m^-,\alpha,\alpha^+,\alpha^-: m\in {\cal{M}},\alpha\in{\cal{A}}\}$,
$p_n^{G}$  and $p_n^{G}(K)$ are respectively defined to be the number of $n$-edge
 $G$-SAPs and $G(K)$-SAPs.
First note that all the after-strand-passage polygons formed
from SAPs counted in $p_m^{m}$ or $p_m^{m^+}$, $m \in\{14,16,18,20,22\}$, are unknotted.
\ Also note that SAPs counted in $p_{14}^{14}$ either contain juxtaposition $Z$ or its mirror,
$\widetilde{Z}$; while those counted in $p_{14}^{14^+}$ all contain
juxtaposition $Z^+$.

One goal is to investigate the asymptotic (as $n\to\infty$) properties of the knot probabilities $\rho_n^G(K)$ for  knot-type $K\in{\cal{K}}$ and geometry $G$.  Towards this end, the asymptotic properties of $\rho_n^G(K)$'s numerator and denominator polygon counts are explored first.  In the next section we prove that both these terms grow exponentially at the same rate. We then make conjectures (based on heuristic arguments) about the asymptotic properties of $\rho_n^G(K)$.

\subsection{Asymptotic properties of {$\rho_n^G(K)$}}
\label{asymsec}

For the set of all SAPs in $\mathbb{Z}^3$, Sumners and Whittington (1989) \cite{Sum88} proved that the following limit exists:
\begin{equation}
\lim_{n\to\infty} n^{-1}\log p_n(\phi)=\lim_{n\to\infty} n^{-1}\log u_n(\phi)=\kappa_0\end{equation} and satisfies \begin{equation} \kappa_0 < \kappa:= \lim_{n\to\infty} n^{-1}\log u_n,
\end{equation}
where $u_n$ is the total number (up-to-translation) of $n$-edge SAPs in the simple cubic lattice and $u_n(\phi)$ is the number (up-to-translation) of these that are unknotted.  The following estimates for $\kappa$ are available
$\kappa=1.544148 \pm
0.000034$ \cite{rechbuks} (via Monte Carlo)  and  $1.544162\pm 0.000219$ \cite{cls07} (via exact enumeration) \cite{cls07}.  For $\kappa_0$, a recent estimate for the difference $\kappa-\kappa_{0}$ is $
(4.15\pm0.32)\times10^{-6}$ \cite{j02}.
The next order behaviour for these polygon counts is not known rigorously but it is widely believed \cite{otjw98}, backed up by numerical evidence \cite{cls07,otjw96,otjw98,ojtw98}, that there exist real numbers $A_0$ and $\alpha_0$ such that, as $n\to\infty$:
\begin{equation}
\label{asymform}
p_n(\phi)=A_0n^{\alpha_0}e^{\kappa_0n}(1+o(1)).
\end{equation}

To explore the asymptotic properties of $\rho_n^{\Theta_0}(K)$, we focus first on $p_n^{\Theta}$, $p_n^{\Theta_0}(K)$, and $p_n^{\Theta_0}$, and establish relationships between them and $p_n(\phi)$. First note that every $n$-edge $\Theta$-SAP is an $n$-edge unknotted SAP rooted at the origin and therefore
\begin{equation}
p_n^{\Theta_0}(K)\leq p_n^{\Theta_0}\leq p_n^{\Theta}\leq p_n(\phi)=nu_n(\phi).
\end{equation}

\begin{figure}[hhh] \psfrag{x}{\tiny{$x$}}
\psfrag{y}{\tiny{$y$}}\psfrag{z}{\tiny{$z$}}
\centering{\includegraphics[
height=2.5857in,
width=4.4149in
]
{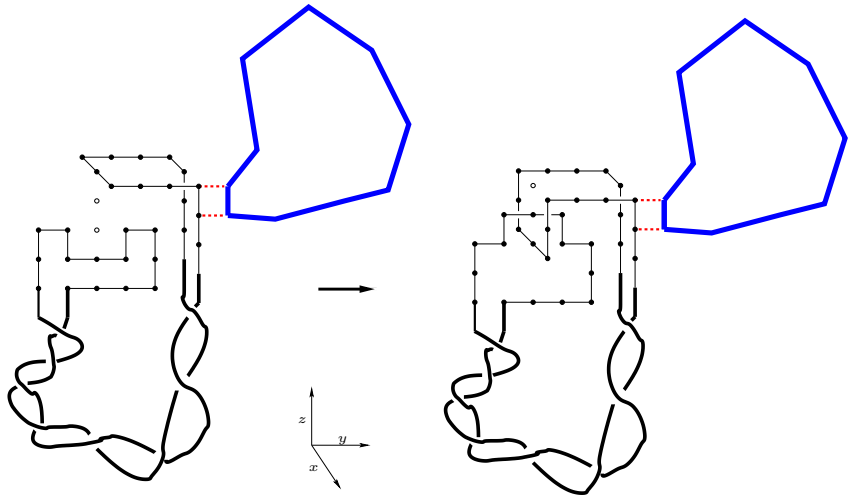}}
\caption{On the left, a schematic depiction of the concatenation of an arbitrary unknotted SAP to a $\Theta_0$-SAP;  on the right, the resulting knot-type $10_1$ after-strand-passage polygon.}\label{fig_twist_SAP}
\end{figure}

Next we show that, given any knot-type $K\in{\cal{K}}$,  there exists an integer $m_K$ such that for any $n\geq m_K$, $u_{n-m_K}(\phi)/2\leq p_n^{\Theta_0}(K)$. To do this,
given any $K\in{\cal{K}}$ and any sufficiently large integer $n$, we  present a method for constructing an element counted in $p_n^{\Theta_0}(K)$ (an $n$-edge $\Theta_0(K)$-SAP) from an unknotted SAP.  The construction will involve two steps.
The first step is the construction of a specific  $\Theta_0(K)$-SAP, $\omega_K$, with length $m_K$.  Then, for the second step, given any $(n-m_K)$-edge unknotted SAP $\omega$, a $\Theta_0(K)$-SAP of length $n$ is constructed by ``concatenating'' $\omega$ to $\omega_K$. The latter step involves defining a way to concatenate a SAP to a $\Theta_0$-SAP so
that the result is still a $\Theta_0$-SAP. (See figure \ref{fig_twist_SAP} for a schematic description of this argument for the case
that $K=10_1$.)  The full details of both steps are given next.  Note that this argument was first presented in \cite{s09} but we expand on the details  here in order to illustrate that the argument is more widely applicable.

For the first step of the construction, in the case that $K=\phi$ (the unknot),
figure \ref{graph_defn_theta} (c) shows a 14-edge $\Theta_0(\phi)$-SAP.  Thus we can take $\omega_{\phi}$ to be the
polygon in figure \ref{graph_defn_theta} (c) and set $m_{\phi}=14$, the number of
edges in $\omega_{\phi}$.
For any other $K$, i.e. $K\in{\cal{K}}\setminus\{\phi\}$, since $K$ has unknotting number one, by definition \cite{m04}, there exists a knot diagram (ie a signed projection into $\mathbb{R}^2$) of it and a crossing, $X$, in the diagram such that: when the sign assigned to $X$ is changed, the result is the knot diagram $D$ of an unknot.  For a fixed such diagram $D$ and the corresponding crossing $X$, $\omega_K$ is formed as follows.
First, deform and subdivide $D$ so that it gives a signed embedding, $D'$, in $\mathbb{Z}^2$ such that signs are assigned to vertices of degree 4 and the signed vertices are in one-to-one correspondence with the signed crossings in $D$.   Respecting the signs of the vertices in $D'$,  add vertical  (and, as needed, planar) edges and then translate and rotate to produce from $D'$ an unknotted $\Theta_0$-SAP where the vertices $B$ and
$F$ of $\Theta$ correspond to the crossing $X$ in $D'$.  The result will
be a $\Theta_0(K)$-SAP and hence such $\Theta_0$-SAPs exist. Let $\omega_K$ be a $\Theta_0(K)$-SAP with the least possible number of edges and let $m_K$ be the number of edges in $\omega_K$.   (Note that in this construction the crossing sign of $\Theta$  is fixed to be consistent with the crossing $X$; hence, depending on the chirality of $K$, the same argument can be applied to construct either a $\Theta^+$- or a $\Theta^-$-SAP or a $J^+$- or $J^-$-SAP.)

For the second step, we recall that the standard (for precise details cf.  \cite[section 1.2.1]{buksbook} or  \cite[algorithm 2.2.2]{s09}) procedure for concatenation of polygon $\omega_2$ to $\omega_1$ in $\mathbb{Z}^3$ involves: translating $\omega_2$ so that its bottom-most edge, $e_2$, is one unit in the positive $x$ direction from the top-most edge, $e_1$, of $\omega_1$; then (if necessary) rotating
$\omega_2$ around the $x$-axis so $e_2$ becomes parallel to
$e_1$; and finally deleting $e_1$ and $e_2$ and then
joining the two polygons by adding in two new parallel edges in the positive $x$ direction. It is straightforward to show from the definitions (cf. \cite[corollary 2.2.3]{s09}) that the standard concatenation of any unknotted SAP $\omega_2$ to any $\Theta$-SAP $\omega_1$ yields a
$\Theta$-SAP.   Thus, for any even $n\geq m_K+4$, concatenating any $n-m_K$
edge unknotted SAP $\omega$ to $\omega_K$ yields an $n$-edge  $\Theta_0(K)$-SAP.  Note that since there are only two possible choices for the initial orientation
of $e_2$ relative to $e_1$, ie either they are parallel or perpendicular,
then at most two different choices of $\omega_2$ could lead, via this concatenation procedure, to the same concatenated polygon.

Since $K\in{\cal{K}}$ was chosen arbitrarily and $\omega_2$ was an arbitrary $(n-m_K)$-edge unknotted SAP in the above construction, we conclude:
for each $K\in{\cal{K}}$, there exists an integer $m_K\geq 14$ such that for any even $n\geq m_K+4$
\begin{equation}
\frac{1}{2}u_{n-m_K}(\phi)\leq p_n^{\Theta_0}(K)\leq p_n^{\Theta_0}\leq p_n^{\Theta}\leq p_n(\phi),
\label{maininequality}
\end{equation}
where the factor of $1/2$ accounts for the possibility that two
different $\omega_2$ yield identical concatenated polygons.
Hence
\begin{equation}
\kappa_0=\lim_{n\to\infty} n^{-1}\log p_n^{\Theta_0}(K)=\lim_{n\to\infty} n^{-1}\log p_n^{\Theta_0} = \lim_{n\to\infty} n^{-1}\log p_n^{\Theta}. \label{eq10}
\end{equation}

As noted above, the construction leading to equation (\ref{maininequality}) will apply also to
any juxtaposition $J^+$ or $J^-$, $J\in{\cal J}$, and  $K\in{\cal{K}}$ that can result
from performing strand passage at such a juxtaposition.     Thus,  given any geometric
specification $G$ which is defined in terms of either a single signed juxtaposition or any group
of signed juxtapositions, we have also that:
for each $K\in{\cal K}$, there exists an integer $m_K^{G}\geq 14$ such that for any even $n\geq m_K^{G}+4$
\begin{equation}
\frac{1}{2}u_{n-m_K^G}(\phi)\leq p_n^{G}(K)\leq p_n^{\Theta_0}\leq p_n^{\Theta}\leq p_n(\phi).
\label{maininequality2}
\end{equation}
Hence, for example,
\begin{equation}
\kappa_0=\lim_{n\to\infty} n^{-1}\log p_n^{G}(K)=\lim_{n\to\infty} n^{-1}\log p_n^{\Theta_0} = \lim_{n\to\infty} n^{-1}\log p_n^{\Theta}. \label{eq102}
\end{equation}

One direct consequence of this result with respect to the asymptotic properties of the knot probabilities is that $\rho_n^{\Theta_0}(K)$ and $\rho_n^G(K)$ do not grow (or decay) exponentially with $n$, that is:
\begin{equation}
\label{eq11}
\lim_{n\to\infty} n^{-1} \log \rho_n^{\Theta_0}(K) =\lim_{n\to\infty} n^{-1}\log \rho_n^G(K)=0.
\end{equation}

Another consequence is that, as $n\to\infty$, each of $p_n(\phi)$, $p_n^{\Theta}$, $p_n^{\Theta_0}$, $p_n^{\Theta_0}(K)$, $p_n^G$ and $p_n^G(K)$  can be written in the form $e^{\kappa_o n +o(n)}$.
It is expected that, just like for $p_n(\phi)$, the more detailed asymptotic behaviour of each of these quantities has a form similar to that
given in equation (\ref{asymform}) but where the {\it amplitude} $A_0$ and
{\it critical exponent} $\alpha_0$ of equation (\ref{asymform}) may be quantity dependent.  For convenience, we use the notation $A_{*}$ and $\alpha_{*}$ to denote, respectively, the amplitude and critical exponent corresponding to the
polygon counts for $n$-edge $*$-SAPs for each $*\in\{G,G(K):G\in{\cal{G}},K\in{\cal{K}}\}$ with ${\cal{G}}=\{\Theta,\Theta_0,\Theta^{\sigma},\Theta_0^\sigma,J,J^\sigma,m,m^\sigma,\alpha,\alpha^\sigma:J\in{\cal{J}},\sigma\in\{+,-\},m\in{\cal{M}},\alpha\in{\cal{A}}\}$. The existence of the limits that
would define these amplitudes or critical exponents has not been proved for any of these quantities. Instead, we next give a heuristic argument that
leads us to make conjectures about relationships between the critical exponents and about the asymptotic behaviour of $\rho_n^{\Theta_0}(K)$.  These conjectures are then
investigated numerically in sections~\ref{mlesec} and \ref{PROBABILITY_SECTION}.

Consider the $\Theta_0$-SAP $\omega_{\phi}$ shown in figure \ref{graph_defn_theta} (c).  Removing
any one of the polygon edges which is not part of $\Theta$
yields a ``pattern'', $P_{\phi}$, which can occur as a subwalk of some unknotted polygon more than once and indeed arbitrarily often.  Similarly such a pattern $P_K$ can be obtained from $\omega_K$ for each $K\in {\cal K}\setminus\{ \phi\}$.  Consistent with what is known for all polygons \cite{Sum88}, it is believed (although not proved) that, given one of these patterns $P_K$, there exists an $\epsilon >0$ such that  all but exponentially few sufficiently large $n$-edge unknotted SAPs contain $P_K$
as a subwalk at least $\epsilon n$ times. If this were true, then almost all
large enough unknotted polygons contain several copies of  $P_K$ and hence
several copies of translated versions of $\Theta_0$.  Such an unknotted polygon
can then be translated so that any one of the $\epsilon n$ $P_K$'s contains $\Theta_0$ and
the resulting polygon is a distinct $\Theta_0(K)$-SAP.  This leads us to
the following conjecture.

\begin{conj}
Given any $K\in{\cal{K}}$, there exists
${\epsilon}_K>0$ and integer $N_K>0$ such that for all $n\geq N_K$:
\begin{equation}
{{\epsilon}_K n}u_{n}(\phi)\leq p_n^{\Theta_0}(K).
\end{equation}
\label{conj1}
\end{conj}

If the critical exponents exist and if conjecture \ref{conj1} is true
then the following is a direct consequence.

\begin{conj}
$\alpha_{\Theta}=\alpha_{\Theta_0}=\alpha_{\Theta_0(K)}
=\alpha_0$
for all  $K\in{\cal{K}}$.
\label{conj2}
\end{conj}

Furthermore if the asymptotic form of equation (\ref{asymform}) applies
generally then the following conjectures are a further consequence.

\begin{conj}
Given any $K\in{\cal{K}}$,
\begin{equation}0<\rho^{\Theta_0}(K):=\lim_{n\to\infty} \rho_n^{\Theta_0}(K)= \frac{A_{\Theta_0(K)}}{A_{\Theta_0}} <1.\end{equation}
\label{conj3}
\end{conj}

\begin{conj}
Given any knot-type $K\in{\cal{K}}$ and any $G\in{\cal G}$,
\begin{equation}0<\rho^G(K):=\lim_{n\to\infty} \rho_n^G(K)= \frac{A_{G(K)}}{A_{G}} <1.\end{equation}
\label{conj4}
\end{conj}

Equation (\ref{eq11}) and conjectures \ref{conj2} and \ref{conj3} 
will be investigated numerically, based on Monte Carlo data, in sections \ref{mlesec} and \ref{PROBABILITY_SECTION} respectively. 
Conjecture \ref{conj4} will be explored in  section \ref{juxtasec} with $G$ ranging from specific
juxtapositions to groupings such as the compactness-size grouping or the opening angle grouping.  The details of the Monte Carlo simulations used to generate the data for these studies is presented next.

\section{Simulation Specifics}
\label{thetabfacfsection}
To study $\Theta$-SAPs, Szafron \cite{s00} developed the $\Theta${-BFACF algorithm} by modifying the BFACF algorithm~\cite{bf81,cc83,ccf83} to only generate $\Theta$-SAPs.
Based on the arguments of Janse van Rensburg and Whittington~\cite{jw91}, Szafron \cite{s00}  proved that the $\Theta${-BFACF algorithm} preserves the knot-type
of the initial polygon. This non-trivial proof relies on the fact that the two strands of the structure $\Theta$ have enough lattice space separating them to  allow another polygon strand  to pass through; this guarantees that Reidemeister III moves are possible
using a sequence of BFACF moves.  In the case that the initial polygon is unknotted, Szafron proved that the
$\Theta${-BFACF algorithm} has two ergodicity classes which correspond precisely to the $\Theta^+-$ and $\Theta^-$-SAPs defined in the previous section.
However, as seen from equations \ref{connectionclassrelation}-\ref{juxtconn}, all the quantities of interest can be investigated using, for example, $\Theta^-$-SAPs only, and we focus on these.   Thus, from a run of the 
$\Theta$-BFACF algorithm, a sample of $\Theta^-$-SAPs is obtained.
Estimates of the one-step transition knot probabilities of interest can be then obtained from such a sample by performing a single strand passage on each sample polygon  and recording the knot-types for each of the resulting after-strand-passage polygons.

The $\Theta$-BFACF algorithm generates a Markov Chain $\{X_{t},t=0,..,T\}$
such that at each time $t,$ $X_{t}$ is a $\Theta$-SAP in the same ergodicity class (crossing-sign-class) as $X_0$. Here $X_0$ is an unknotted $\Theta^-$-SAP and we define the set of these to be $\mathscr{P}$. The standard three possible BFACF moves are used to go from $X_{t}$ to $X_{t+1}$, however, the probability distribution
according to which a move is attempted is modified from that of the BFACF algorithm to accommodate for the reduced state space.

For a fixed integer
$q$ satisfying $z_{\phi}^2:=e^{-2\kappa_0} \leq \left[\frac{2}{3} \right]^{q-1}$ and a fixed real-valued $z$ such that $0<z<z_{\phi},$ the $\Theta$-BFACF algorithm one-step transition probabilities $P_{\omega\omega^{\prime}}=P(X_{t+1}=\omega^{\prime}|X_t=\omega)$, for all $\omega,\omega'\in\mathscr{P},$
are chosen so that the
equilibrium probability distribution, $\{\pi_{\omega}(q,z),\omega\in {\mathscr{P}}\}$, of the Markov Chain 
is given by
\begin{equation}
\pi_{\omega}(q,z)=\frac{(|\omega|-6)|\omega|^{q-1}z^{|\omega|}}{\sum_{i=0}^{\infty}(i-6)i^{q-1}p_{i}^{\Theta}z^{i}}, ~\mbox{\rm for all }~\omega\in\mathscr{P}.\label{eqn_eq_dist_BFACF}
\end{equation}

Because the $\Theta$-BFACF algorithm is based on the BFACF algorithm,
it suffers from the same major disadvantage, that is, as $z\rightarrow z_\phi$, the
exponential autocorrelation time for the algorithm approaches infinity \cite{st89}.
\ Hence, to reduce the exponential autocorrelation time, a composite
Markov chain (CMC) implementation of the $\Theta$-BFACF algorithm is used.
It is similar to the multiple Markov chain BFACF algorithm introduced in \cite{otjw96,ojtw98} except now the equilibrium distribution
for a single chain is given by equation (\ref{eqn_eq_dist_BFACF}).

Given any integer $M\geq 1$ and a real vector $\boldsymbol{z}:=(z_1,z_2,...,z_M)$ such that $0<z_1<z_2<...<z_M<z_\phi$, Szafron \cite{s00} proved that the CMC  $\Theta$-BFACF
algorithm  is ergodic
on $[\mathscr{P}]^{M}$ and has the unique stationary distribution given by
\begin{equation}
\{{\pi}_{\boldsymbol{\omega}}(q,\boldsymbol{z}),{\boldsymbol{\omega}
\in\mbox{$\mathscr{P}$}^{M}}\}, \label{CMCpmf}
\end{equation}
where for $\boldsymbol{\omega}=(\omega_1,\omega_2,...,\omega_M)\in\mathscr{P}^{M}$,
\begin{equation}
{\pi}_{\boldsymbol{\omega}}(q,\boldsymbol{z}):=\prod_{i=1}^M  {\pi
}_{\omega_i}(q,z_{i}), \label{CMCpmfb}
\end{equation}
with ${\pi
}_{\omega_i}(q,z_{i})$, the distribution of the $i$th chain,   as in equation (\ref{eqn_eq_dist_BFACF}).

The simulation of the CMC $\Theta$-BFACF algorithm used for this work consisted
of ten independent replications. For the results in \cite{s09} and section \ref{mlesec}, each replication was run for a total
of $1.8\times10^{11}$ time steps ($1.5\times10^{11}$ $\Theta$-BFACF
moves in parallel and $0.3\times10^{11}$ attempted swaps) where every
sequence of five $\Theta$-BFACF moves in parallel was followed by an attempted
swap between a randomly selected chain (call it chain $i$) and chain $i+1$.
While for the results in sections \ref{PROBABILITY_SECTION} and \ref{juxtasec},
each replication was extended to 250 billion $\Theta$-BFACF moves in parallel.
For the  distribution given
by equation (\ref{eqn_eq_dist_BFACF}), $q$ is set to $2$. \ For each
 individual replication, the number of chains and the distribution
of the $z_{i}$'s over the interval $[0.2030,0.2132]$  is: $M=14$, $z_{1}=0.2030$, $z_{2}=0.2050$, $z_{3}=0.2070$, $z_{4}=0.2090$,
$z_{5}=0.2100$, $z_{6}=0.2105$, $z_{7}=0.2110$, $z_{8}=0.2115$,
$z_{9}=0.2120$, $z_{10}=0.2124$, $z_{11}=0.2128$, $z_{12}=0.2130$,
$z_{13}=0.2131$, and $z_{14}=0.2132$. These values of $z$ are valid
for the $\Theta$-BFACF algorithm because for $i=1,\ldots,14$,
$$
z_{i}<z_{\phi}<0.2135$$
\cite{g87,s00}. One motivation for using this distribution
of $z$-values and $M=14$ is that these choices have been well studied for unknotted SAPs using the BFACF algorithm (cf. \cite{otjw96,otjw98}).

The amount of time
required for the entire process to equilibrate $(\tau_{\exp})$ was estimated using Gelman and Rubin's Estimated Potential Scale Reduction technique
\cite{g96,gr92}.
Applying this technique, 
we estimate $\widehat{\tau}_{\exp}=5.0$ billion
$\Theta$-BFACF moves in parallel (i.e. after $5.0$ billion $\Theta$-BFACF
moves in parallel, the estimated between-the-replication and within-a-replication
variances have converged to within 2.5\% of the same value). 

In order to estimate the amount of ``essentially independent''
data collected during each replication, Fishman's Block Analysis (cf. \cite{f97}) technique
was used.
This  technique
yielded $\widehat{\tau}_{{int}}=0.7$ billion $\Theta$-BFACF
moves in parallel and hence we conclude that states that are  $1.4$ billion $\Theta$-BFACF
moves in parallel apart (ie $2\widehat{\tau}_{{int}}$ apart) are essentially independent and data that
is subdivided into blocks of 1.4 million consecutive data points form
essentially independent blocks of data. Thus the results presented in section \ref{mlesec} are based on 1070 essentially independent blocks and those in sections \ref{PROBABILITY_SECTION} and \ref{juxtasec}, on 1785 essentially independent blocks.  Sokal
\cite{st89} has argued that, if  $\tau_{exp}$  is
less than 5\% of the total length of the replication, then the bias introduced
into the estimates by not discarding the first $\tau_{exp}$ data points will be
much smaller than the actual statistical error. \ Based on this argument, no data was discarded for any of our estimates (40 blocks is less than 5\% of  $1070<1785$  blocks). 

\section{Maximum Likelihood Estimates and Limiting Knot Probability Results}

We have developed \cite{s09} two methods for statistical analysis of the CMC $\Theta$-BFACF Monte Carlo data.  The first is a method for obtaining maximum likelihood estimates for the growth constants and critical exponents of the polygon counts involved in the knotting probability calculation.
The second is a method for investigating the length ($n$) dependence of the knotting probabilities using a correlated sequence of polygon data from a CMC run; the effect of the correlation is reduced by grouping polygons having lengths within a given range to obtain what we call {\it grouped-$n$} estimates.  
Next, in section \ref{mlesec} and appendix A, we summarize the 
maximum likelihood method and present results related to equation (\ref{eq11}) and conjecture \ref{conj2}. In section \ref{PROBABILITY_SECTION}, we review the grouped-$n$ estimate approach and present results related to conjecture \ref{conj3} using data beyond that of \cite{s09}.

\subsection{Maximum Likelihood Estimates from CMC \boldmath{$\Theta$}-BFACF Data}
\label{mlesec}In \cite{bs85}, a method (referred to here as the Berretti-Sokal
MLE Method) was proposed for obtaining maximum likelihood
estimates (MLEs) for $\kappa$ and $\gamma$ (where $\kappa$ and $\gamma$
are exponents in the asymptotic form for the number, $c_n$,  of $n$-step self-avoiding walks (SAWs) starting at the origin, that is $c_{n}\sim Ae^{\kappa n}n^{\gamma-1}$) from
a Markov Chain Monte Carlo simulation consisting of several independent
sample paths. In \cite{s09},  the Berretti-Sokal
MLE method was modified for the case that the Markov Chain data comes from a
CMC Monte Carlo sample path.  For completeness, we summarize the new MLE method of \cite{s09} in appendix A and review in this section the results regarding the asymptotic properties of the $\Theta$-SAP counts.

The main purpose of the MLE method is to investigate
conjecture \ref{conj2} and also to obtain an estimate of
$\kappa_0$ (see equation (\ref{eq10})).
To explore conjecture \ref{conj2},
statistical estimates of $\alpha_{*}$, $*\in{\cal G}$, are needed.  To do this, for a given choice of $*$, a log-likelihood
function is obtained based on the CMC Monte Carlo data generated.  Maximizing the log-likelihood with respect to $\kappa_0$ and $\alpha_{*}$ results in maximum likelihood estimates for these parameters.
The relevant log-likelihood function is defined in appendix A.

Using this  method
yields the following estimates for $\kappa_0$ (cf. table \ref{table_kappa_MLEs}) and $\alpha_{*}$ where $*\in \{\Theta_0,\Theta_0(\phi),\Theta_0(3_1)\}$ (cf. table~\ref{table_alpha_MLEs}).
\begin{table}[tttt]
 \centering
\caption{Our CMC MLEs for $\kappa_{0}$.
The values in parentheses are the estimated $95\%$ margins of error.}
\begin{tabular}{|l|l|l|l|}
\hline
 & \multicolumn{3}{|c|}{Parameter Estimated}\tabularnewline
\hline
Property $\ast$  & $N_{\min}^{\ast}$  & $N_{\max}^{\ast}$  & $\kappa_{0}$ $(95\%$ ME)\\
\hline
$\Theta_0$  & 98  & 3300  & $1.544148\left(0.000012\right)$\tabularnewline
\hline
$\Theta_0\left(\phi\right)$  & 86  & 3300  & $1.544147\left(0.000013\right)$\tabularnewline
\hline
$\Theta_0\left(3_{1}\right)$  & 162  & 2000  & $1.544135\left(0.000023\right)$\tabularnewline
\hline
\end{tabular}\label{table_kappa_MLEs}
\end{table}
The estimates for $\kappa_{0}$ 
in table \ref{table_kappa_MLEs} are all equal to four decimal places
and are equal (after rounding) to four decimal places to a previous direct estimate:  $\kappa_{0}=1.544067 \pm 0.000811$ \cite{otjw98}.
Thus our estimates
for $\kappa_{0}$ numerically support the proven result, equation~(\ref{eq10}), that $p^{\Theta_0}_n$ and $p_n^{\Theta_0}(K)$ grow at the same exponential rate as $p_n(\phi)$ as $n\to\infty$.

Because it provides our largest data sample,  we use all the sampled $\Theta^-$-SAPs in the MLE analysis  to determine our
best estimates for $\kappa_{0}$  and $\alpha_{\Theta}$.
Our resulting best estimates for $\kappa_{0}$ and $\alpha_{\Theta}$
are:
\begin{equation}
\kappa_{0}=1.544148\pm0.000014\left(\pm0.00005\right)
\end{equation}
and
\begin{equation}
\alpha_{\Theta}=-1.78\pm0.02\left(\pm0.02\right), \label{95_conf_int_alpha_phi}
\end{equation}
given in the form
\begin{equation}
\nonumber
\mbox{\rm parameter}=\mbox{\rm point estimate}\pm95\%\ \mbox{\rm ME }(\pm\mbox{\rm systematic error}).
\end{equation}
With respect to the systematic error term, we report an error associated with  the uncertainty in the choice of $N_{\min}$ (defined in the appendix) which is obtained by taking the largest difference (over a range of $N_{\min}$ choices) between the resulting estimates and our reported point estimate.

Another estimate for $\kappa_0$ is $1.544158$, which is based on the estimate for $\kappa \approx 1.544162\pm 0.000219$ (Clisby \emph{et al.} \cite{cls07})  and the estimate for the difference $\kappa-\kappa_{0}$$\approx
(4.15\pm0.32)\times10^{-6}$ (Janse van Rensburg \cite{j02}).  Our best estimate for $\kappa_0$ is  consistent with this.
Our best estimate for $\alpha_{\Theta}$ is used next to explore the validity of
conjecture~\ref{conj2}.

From table \ref{table_alpha_MLEs},
the estimates for $\alpha_{\Theta}$, $\alpha_{\Theta_0}$ and $\alpha_{\Theta_0(\phi)}$ are equal when rounded to two decimal places; this supports that $\alpha_{\Theta}=\alpha_{\Theta_0}=\alpha_{\Theta_0(\phi)}$ as in conjecture~\ref{conj2}.  The computed $95\%$ confidence interval for $\alpha_{\Theta_0(3_1)}$ completely contains the computed $95\%$ confidence intervals for $\alpha_{\Theta_0}$ and $\alpha_{\Theta_0(\phi)}$; this does not contradict the conjecture (part of conjecture~\ref{conj2}) that $\alpha_{\Theta_0(3_1)}=\alpha_{\Theta_0}=\alpha_{\Theta_0(\phi)}$.

\begin{table}[tttt]
 \centering
\caption{The best CMC MLEs for $\alpha_{\Theta_0}$, $\alpha_{\Theta_0(\phi)}$
and $\alpha_{\Theta_0(3_1)}$.
The values in parentheses are the estimated $95\%$ margins of error.}
\begin{tabular}{|l|l|l|l}
\hline
 & \multicolumn{3}{|c|}{Parameter Estimated}\tabularnewline
\hline
Property $\ast$  & $N_{\min}^{\ast}$  & $N_{\max}^{\ast}$  &  \multicolumn{1}{l|}{$\alpha_{\ast}$ $(95\%$ ME)}\tabularnewline
\hline
$\Theta_0$  & 98  & 3300  &  \multicolumn{1}{|r|}{$-1.7804\left(0.0304\right)$}\tabularnewline
\hline
$\Theta_0\left(\phi\right)$  & 86  & 3300  & \multicolumn{1}{|r|}{$-1.7793\left(0.0229\right)$}\tabularnewline
\hline
$\Theta_0\left(3_{1}\right)$  & 162  & 2000  & \multicolumn{1}{|r|}{$-1.9498\left(0.2938\right)$}\tabularnewline
\hline
\end{tabular}\label{table_alpha_MLEs}
\end{table}

To explore the last part of conjecture~\ref{conj2}, that all the $\alpha$'s are
equal to $\alpha_0$, we use our best estimate for the critical exponent
$\alpha_{\Theta}$ (assumed to be equal to $\alpha_{\Theta_0}=\alpha_{\Theta_0(K)}$ (for any  $K\in{\cal K}$)).
In order to compare this to $\alpha_{0}$, note that
Orlandini \textit{et al}~\cite{otjw98} estimated $\alpha_{0}-1$
$\approx-2.77$. \ Using this value for $\alpha_{0},$ gives $\alpha_{0}=-1.77$. \ Since this value is contained
in our estimated 95\% confidence interval for $\alpha_{\Theta}$
given by equation (\ref{95_conf_int_alpha_phi}), the final part of conjecture~\ref{conj2}
is supported numerically$.$

\subsection{Estimating the Limiting Knot Probabilities}
\label{PROBABILITY_SECTION}In this section, we explore conjecture~\ref{conj3}.
The quantities $\rho^{\Theta_0}$ and
$\rho^{\Theta_0}(K)$, for each $K\in\mathcal{K}$, will generally be referred to as
\textit{limiting probabilities}.
As discussed in section \ref{asymsec}, the existence of these limiting probabilities is an open
question. Conjecture~\ref{conj3} postulates that, not only do these limits exist, but that they are never zero or one.

To obtain an indication of the order of magnitude of these limiting probabilities,
the frequencies of $\Theta^-_0(K)$-SAPs observed in our CMC data are summarized in table~\ref{totalobservedpolygonspropstar}.
(Note that the ``Other'' category in table~\ref{totalobservedpolygonspropstar} contains the observed number of $\Theta^-_0(K)$-SAPs over all $K\in{\cal K}$ with crossing number greater than five.)  Thus, roughly, we expect the limiting probabilities for the: unknot to be close to $1$; trefoil to be of the order $10^{-2}$; figure eight to be of the order $10^{-4}$; knot-type $5_2$ to be of the order $10^{-5}$;  and, at least a six-crossing knot to be of the order $10^{-7}$.  This ranking and the orders of magnitude for the $\phi\to K$ knot probabilities is comparable to that obtained for other strand-passage models.  For example,
in \cite[column 1 of Table 2]{hnrav07} for $K=\phi, 3_1, 4_1, 5_2$ the probabilities (for lattice polygons with mean length 100) are
0.852, 0.061, 0.022, 0.0016 and in \cite[column 1 of Table 1]{fms04}  
(for freely jointed isolateral polygons of length 33) they are 0.9457, 0.0227, 0.0073, 00006.  We expect these probabilities to be both length and model dependent and hence we do not make more direct comparisons between the $\Theta$-SAP model and these other models.
Note that for each $K\in\cal{K}$ with $K$ being chiral, the after-strand-passage polygons we observed were all in the same chirality class, that is, for example, we only observe $3_1^+$ after-strand-passage trefoils and $5_2^+$ after-strand-passage five crossing knots.
\begin{table}
\centering\begin{tabular}{|l|c|}
\hline
Property $\ast$  & Frequency\tabularnewline
\hline
$\Theta^-_0$  & \multicolumn{1}{|r|}{2491776147}\tabularnewline
\hline
$\Theta^-_0(\phi)$  & \multicolumn{1}{|r|}{2459748925}\tabularnewline
\hline
$\Theta^-_0(3_{1})$  & \multicolumn{1}{|r|}{31161421}\tabularnewline
\hline
$\Theta^-_0(4_{1})$  & \multicolumn{1}{|r|}{828162}\tabularnewline
\hline
$\Theta^-_0(5_{2})$  & \multicolumn{1}{|r|}{36596}\tabularnewline
\hline
Other  & \multicolumn{1}{|r|}{1029}\tabularnewline
\hline
\end{tabular}\caption{The number of polygons observed across the ten replications
and the fourteen chains that have property $\ast$.}
\label{totalobservedpolygonspropstar}
\end{table}

Although the frequencies presented in table~\ref{totalobservedpolygonspropstar} can be used to estimate the approximate order of magnitude of $\rho^{\Theta_0}(\ast)$, they cannot be used to directly estimate $\rho^{\Theta_0}(\ast)$, $K\in{\cal K}$.  
Instead, in order to study conjectures~\ref{conj3} and~\ref{conj4}, we suppose that the polygon counts for $n$-edge $\ast$-SAPs, for $\ast\in \{G,G(K):G\in\cal{G},K\in\cal{K}\}$, have the asymptotic form
\begin{equation} A_{\ast}n^{\alpha_{\ast}}e^{\kappa_{0}n}\left(  1+\frac{B_\ast}{n^{\Delta_\ast}}+O(n^{-1}))\right), \label{equation_scaleform}\end{equation} 
for some constants (independent of $n$) $B_*$ and $\Delta_*>0$.
From this, it can be shown that, for $G(K)$-SAPs where $G\in\cal{G}$, as $n\to\infty$, there exists  other constants $B^G(K)$ and $\Delta^G(K)>0$ such that
\begin{equation}
\rho_n^G(K)\approx\rho^G(K)+B^G(K)n^{-\Delta^G(K)}.\label{scaling_form_fixedn_Ksp}\end{equation}
In general, the variability in any estimates of 
 $\rho_n^G(\ast)$  increases with $n$.  In order to reduce this variability towards obtaining estimates for the limit $\rho^G(K),$ we use ``grouped-$n$'' polygon counts.  Specifically, for positive even integers $n_2>n_{1}$, we focus on polygons whose lengths are in the interval $[n_{1},n_{2})$ and define
the $[n_1,n_2)$-grouped probability (or grouped probability, for short)
\begin{equation}
\rho^G_{n_{1},n_{2}}(K):=\frac{\sum\limits _{n=n_{1}}^{n_{2}-2}\left[p_{n}^{G}\left(K\right)\sum\limits _{i=1}^{M}\frac{w(n)e^{\beta_{i}n}}{{Q}\left(\beta_{i}\right)}\right]}{\sum\limits _{n=n_{1}}^{n_{2}-2}\left[p_{n}^{G}\sum\limits _{i=1}^{M}\frac{w(n)e^{\beta_{i}n}}{{Q}\left(\beta_{i}\right)}\right]},\label{equation_groupedNKsp}\end{equation}
where each sum is taken through even values of $n$, $w(n)=(n-6)n^{q-1}$, $e^{\beta_i}=z_i$, and $Q(\beta_i)$ is the normalizing sum in the denominator of $\pi_{\omega}(q,z_i)$ from equation (\ref{eqn_eq_dist_BFACF}).  

For any given $G\in{\cal G}$ and $K\in{\cal K}$, substituting the scaling form~(\ref{equation_scaleform}) into equation~(\ref{equation_groupedNKsp}), results in, to first order, that there exists $N_{\min}>0$ such that
\begin{equation}
\rho^G_{n_{1},n_{2}}(K)\approx f^G_K(n_1):=\rho^G(K)+m^G(K)n_{1}^{\lambda^G(K)},\label{scaling_form_exp_prob_Ksp}\end{equation}
for all $n_1\geq N_{\min}$, for some constants $m^G(K)$ and $\lambda^G(K)<0$, and with $\rho^G(K)$ as in
conjecture \ref{conj4}. 
Thus, if the limit in conjecture \ref{conj4} exists, the grouped probabilities have the same limit as the non-grouped probabilities. 

\begin{figure}
\centering\psfrag{P}{\hspace{-0.5in}$\widehat{{\rho}}^{\Theta_0}_{n_{1},n_{2}}(K)$}
\psfrag{N}{$n_{1}$}\psfrag{eunknot}{$\phi$}\psfrag{e3_1}{$3_1$}\psfrag{e4_1}{$4_1$}\psfrag{unknot}{\hspace{-.14in}$f^{\Theta_0}_\phi(n_1)$}
\psfrag{3_1}{\hspace{-.27in}$f^{\Theta_0}_{3_1}(n_1)$} \psfrag{4_1}{\hspace{-.27in}$f^{\Theta_0}_{4_1}(n_1)$}
\includegraphics[width=5.1569in,height=3.6118in]{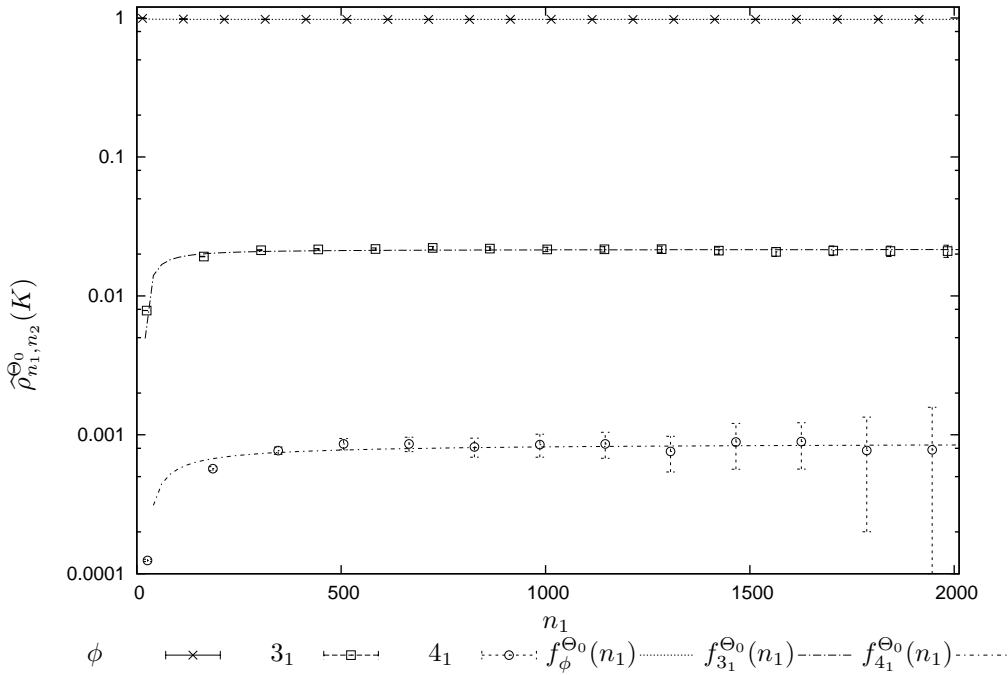}
\caption{The grouped probability estimates $(\times)$ for $\rho^{\Theta_0}_{n_{1},n_{2}}(\phi)$ (for $n_1\in\{14,114,214,...,1914 \}$)  and the fitted curve $(\cdots\cdots)$  $f^{\Theta_0}_{\phi}(n_1)$.  The grouped probability estimates $(\Box)$ for $\rho^{\Theta_0}_{n_{1},n_{2}}(3_1)$ (for $n_1\in\{24,164,304,...,1844 \}$)  and the fitted curve $(-\cdot-\cdot-)$ $f^{\Theta_0}_{3_1}(n_1)$.  The grouped probability estimates $(\circ)$ for $\rho^{\Theta_0}_{n_{1},n_{2}}(4_1|s)$ (for $n_1\in\{30,190,350,...,830 \}$)  and the fitted curve $(-\cdot\cdot-\cdot\cdot-)$ $f^{\Theta_0}_{4_1}(n_1)$. \ The error bars are the estimated
$95\%$ confidence intervals.
\label{fig_prob}}
\end{figure}

For $G\in\cal{G}$, over the interval in which we have reliable data, we determine the non-overlapping intervals $[n_1,n_2)$ in such a
manner that the interval lengths $d^G(K)=|n_2+2-n_1|$ are all constant and that the estimates for
$\sum_{i=n_1}^{n_2-2}p_n^G(K)$ are essentially independent of the estimates for $\sum_{i=n_1+d^G(K)}^{n_2-2+d^G(K)}p_n^G(K)$. 
A procedure for estimating $d^G(K)$ is given in \cite{s09}.
Using this procedure, we estimated that $d^{\Theta_0}(\phi)=100$,
$d^{\Theta_0}(3_1)=140$, and $d^{\Theta_0}(4_1)=160$.  For these choices of $d^G(K)$, the corresponding ratio estimates for the grouped probabilities from our Monte Carlo data are displayed in figure~\ref{fig_prob}, that is figure~\ref{fig_prob}   displays our estimated values for  ${\rho}^{\Theta_0}_{n_{1},n_{2}}(\phi)$ (for $n_1\in\{14,114,214,...,1914 \}$), ${\rho}^{\Theta_0}_{n_{1},n_{2}}(3_{1})$ (for $n_1\in\{24,164,304,...,1844 \}$), and  ${\rho}^{\Theta_0}_{n_{1},n_{2}}(4_{1})$ (for $n_1\in\{30,190,350,...,830 \}$) versus $n_1$.
This figure also displays each of our fitted equations $f_K^G(n_1)$ (from  (\ref{scaling_form_exp_prob_Ksp})) for
 ${\rho}^{\Theta_0}_{n_{1},n_{2}}(\phi)$, ${\rho}^{\Theta_0}_{n_{1},n_{2}}(3_{1})$, and ${\rho}^{\Theta_0}_{n_{1},n_{2}}(4_{1})$
versus $n_{1}$.  

We focus on the sets of observed ${\Theta_0}(\phi)$-, ${\Theta_0}(3_1)$-, and ${\Theta_0}(4_1)$-SAPs because for these SAPs our fitted equation provides a ``good fit'' to  the grouped probability estimates (for sufficiently large $n_1\geq N_{\min}$) over the range of reliable data.  By ``good fit'', we mean that the $p$-value (cf table~\ref{table_prob_est}) associated with a $\chi^2$-Test for Goodness of Fit is larger than 0.05.
Table~\ref{table_prob_est} contains our estimates (from the fit) for the limiting probabilities $\rho^{\Theta_0}(\phi)$, $\rho^{\Theta_0}(3_1)$, and $\rho^{\Theta_0}(4_1)$.  The estimates  have the form
$$\mbox{\rm point estimate}\pm95\%\ \mbox{\rm margin of error}~ (\mbox{\rm systematic error}).$$
The systematic error is estimated by taking the maximum
difference between the grouped probability point estimates over the region
$[14,2012]$ and the corresponding estimated limiting probability; this is a measure for the error resulting from the uncertainty in the choice of $N_{\min}$.  \ \ \

We also analyzed our data for the knot-types with five or more crossings, however, the resulting fits resulted in $p$-values$<0.01$;  hence the corresponding estimated limiting knot probabilities were deemed unreliable.   It should be noted, however, that these estimates did not contradict conjecture~\ref{conj3}.

\begin{table}
\centering
\begin{tabular}{|c|c|c|}
  \hline
 $K$  & $\rho^{\Theta_0}(K)$ & $p$-value \\\hline
 $ \phi$&$ 0.97774 \pm\ 0.00105\ \ (0.005) $&$ 0.42$ \\\hline
  $3_1$& $0.02163 \pm\ 0.00055\ \ (0.003) $& $0.39$ \\\hline
  $4_1$ & $0.00089 \pm\ 0.00018\ (0.0005)$ & $0.17$ \\
  \hline
\end{tabular}
\caption{Estimates for the limiting probabilities $\rho^{\Theta_0}(K)$ determined by fitting an equation of the form $f(x)=b+mx^{r}$ to the estimated grouped probabilities displayed in figure~\ref{fig_prob}.  The estimates are of the form: $\mbox{\rm point estimate}\pm95\%\ \mbox{\rm margin of error}~ (\mbox{\rm systematic error}).$  The $p$-value presented is the $p$-value associated with a $\chi^2$-Test for Goodness of Fit between the grouped probabilities estimated from the data and the grouped probabilities predicted using the fitted equation.}
\label{table_prob_est}
\end{table}

With respect to each of the limiting probabilities $\rho^{\Theta_0}(\phi)$, $\rho^{\Theta_0}(3_1)$, and $\rho^{\Theta_0}(4_1)$, the  corresponding 95$\%$ confidence interval lies completely within the interval $(0,1)$.  Hence we have strong evidence that conjecture~\ref{conj3} holds.

\section{Dependence of the After-Strand-Passage Knotting Probabilities on the Local Juxtaposition}
\label{juxtasec}

In this section we explore how the knotting probabilities depend on the local geometry about $\Theta$ and how they depend on the scheme used to classify the compactness of these local geometries.  We begin by exploring the effect that a small change in the local geometry has on the knotting probability.  Note that, to reduce statistical error, all estimates presented in this section are based on grouped probabilities for grouped polygon-lengths $n=n_1$ to $n_2=n_1+118$.
Also, in this section,  given a specific geometry $G$ and polygon length $n$, the phrase ``knotting probability'' refers to the
$G$-restricted knotting probability, $\rho_n^G(\bar{\phi})$.  

\begin{figure}[ttt]
\centering\psfrag{S1}{\small{$S^-$}} \psfrag{S2}{\hspace{-0.05in}\small{$S^{+}$}} \psfrag{Z1}{\hspace{-0.05in}\small{$Z^-$}}
\psfrag{Z2}{\hspace{-0.05in}\small{$Z^{+}$}}\psfrag{L1}{\hspace{-0.05in}\small{$L^-$}}\psfrag{S}{\hspace{-0.05in}\small{$S$}}
\psfrag{Z}{\hspace{-0.05in}\small{$Z$}} \psfrag{P}{\hspace{-.45in}$\hat{\rho}^{G}_{n_1,n_2}(\bar{\phi})$} \psfrag{N}{$n_1$}
\includegraphics{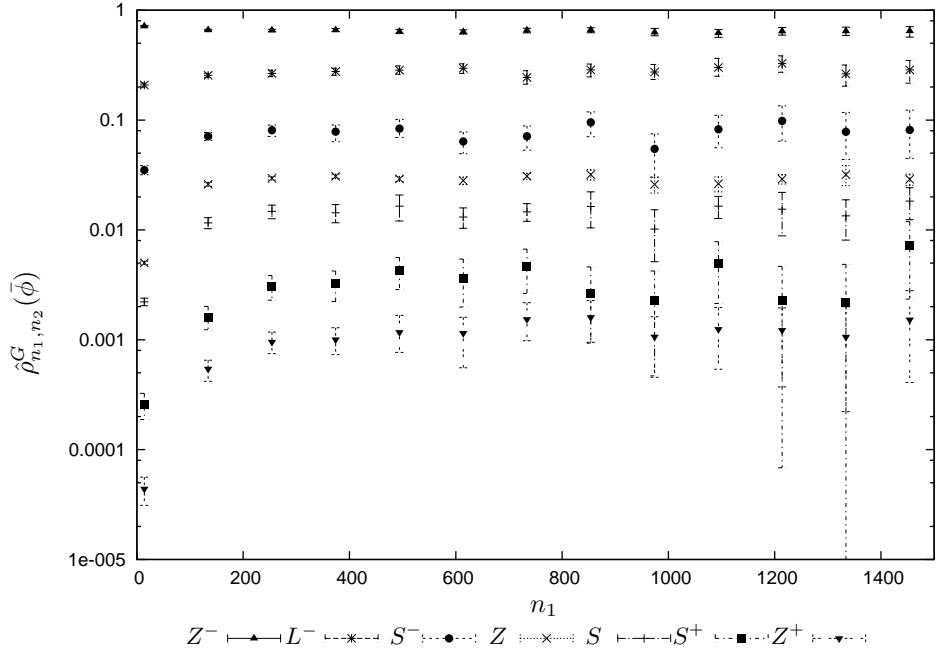}\caption{The grouped-$n$ estimates for the $G$-restricted knotting probabilities (on a log scale) for  $\Theta$-SAPs containing $G=S$, $Z$, the four associated signed juxtapositions, and $L^-$. The error bars are estimated 95\% confidence intervals, $n_1\in\{14,134,...,1454\}$, and $n_2=n_1+118$.}
\label{knotreductvsN}
\end{figure}

To determine the influence that the local geometry can have on  the  knotting probabilities we focus on the three juxtapositions,  $S$, $L$, and $Z$,  illustrated in figure~\ref{free_juxtaposition copy(2)-1-1} because: (1)  the estimated knotting probabilities  for $Z^{-}$ and $Z^{+}$ are, respectively, the highest and lowest  amongst all the juxtapositions, cf. figure~\ref{knotreductvsN}; (2) in our CMC sample, the number of $\Theta_0^-$-SAPs that contain $S$ (17237) is roughly equal to the number of $\Theta_0^-$-SAPs that contain $Z$ (15155); (3) $S$ resembles the  half-hooked juxtaposition of \cite{lmzc06}; and (4) $L$ differs from both $S$ and $Z$ by only one edge.

We first explore the influence of the juxtaposition and the crossing-sign  on the knotting probabilities for juxtapositions $S^{-}$, $S^{+}$, $Z^{-}$, and $Z^{+}$.  
Figure  \ref{knotreductvsN} displays (on a log scale) estimates of the relevant knotting probabilities (along with those for $L^-$ and the unsigned $S$ and $Z$)  versus polygon length
$n_1$.
It is clear from this figure that, for each value of $n_1$, the estimated knotting probabilities for $S^{-}$ and $S^{+}$ and for $Z^{-}$ and $Z^{+}$, respectively, are statistically distinct, with the most dramatic difference being between the estimates for $Z^-$ and $Z^+$.
Now, if  crossing-sign is ignored, then the
associated estimated knotting probabilities (those for the unsigned $S$ and $Z$) in figure \ref{knotreductvsN}  are also statistically distinct for each displayed value of $n_1$, but now the difference is less than that observed, for example, for $S^-$ and $Z^-$. 
Hence the knotting probabilities associated with juxtapositions  can strongly depend on the crossing-sign at the strand passage site.  From conjecture \ref{conj4}, we expect that
the knotting probabilities associated with the juxtapositions (whether signed or unsigned) will go to a juxtaposition-dependent constant as $n\to \infty$.  Figure~\ref{knotreductvsN} provides evidence of this:  for each juxtaposition, the point estimates  appear to be approaching distinct limiting values.
 The estimated limiting knotting probabilities presented in table~\ref{table_95CI_connclass_SLZ} also support this.

We can say more about the influence of the local geometry at the strand passage site.  Although $S$ and $Z$ both differ from $L$ by one edge (cf. figure~\ref{free_juxtaposition copy(2)-1-1}),  for each value of $n_1$, the estimated knotting probabilities for $S^-$, $Z^-$, and $L^-$ (as displayed in figure~\ref{knotreductvsN})  are clearly statistically distinct.  Moreover, the knotting probability for $Z^-$ is approximately double that for $L^-$, and the knotting probability for $S^-$ is approximately one-fifth that for $L^-$.  From the estimates in table~\ref{table_95CI_connclass_SLZ},
the limiting knotting probabilities associated with these three signed juxtapositions  are also statistically different.  Clearly a small change in the local juxtaposition can have a significant impact on the associated probabilities of knotting.
Further, if the crossing-sign dependence is ignored, then the associated estimates for the limiting knotting probabilities (those for $Z$, $L$, and $S$ in table~\ref{table_95CI_connclass_SLZ}) are also statistically different.  We thus conclude that the knotting probabilities are impacted by the local juxtaposition, whether signed or unsigned.

We thus have strong evidence that the crossing-sign  and a very minor change in the local
juxtaposition at the strand passage site  influences the knotting probabilities,  with the most dramatic influence occuring when the crossing-sign is not ignored.  In fact, depending on the juxtaposition geometry  $G$ at the
strand passage site, one can either preferentially knot an unknotted polygon (if $G=Z^-$, for example) or preferentially keep it unknotted (if $G=Z^+$).

\begin{table}
\centering
\begin{tabular}{|c|c|c|c|}
  \hline
  $J$ & $\rho^{J^{-}}(\bar{\phi})$ & $\rho^{J^{+}}(\bar{\phi})$& $\rho^{J}(\bar{\phi})$ \\\hline
  ${Z}$ & $0.629 \pm 0.066$& $0.0013 \pm 0.0002$ & $0.0287 \pm 0.0017$  \\\hline
  ${L}$ & $0.282 \pm 0.036$& $0.0007 \pm 0.0005$ & $0.0223 \pm 0.0052$\\\hline
 ${S}$ & $0.077  \pm  0.012$ & $0.0037  \pm  0.0013$& $0.0153  \pm  0.0042$\\
  \hline
\end{tabular}
\caption{Column 2 displays estimates and 95\% confidence intervals for the limiting knotting probabilities for  $Z^{-}$, $L^{-}$, and $S^{-}$. Column 3 displays estimates and 95\% confidence intervals for the limiting knotting probabilities for  $Z^{+}$, $L^{+}$, and $S^{+}$.  Column 4 displays estimates and 95\% confidence intervals for the limiting knotting probabilities for $Z$, $L$, and $S$.}
\label{table_95CI_connclass_SLZ}\end{table}

We now turn our attention to determining the influence of juxtaposition compactness on the limiting knotting probabilities.  For $m^-$-SAPS, $m\in{\cal M}$, we first study  the dependence of the  proportion of $m^-$-SAPS, $\displaystyle{\frac{p_n^{m^-}}{p_n^{\Theta_0^-}}}$, on $m$.
The second column in table~\ref{table_95CI_propcompclassI} displays 95\% confidence intervals for the limiting proportion of $m^-$-SAPS, $\displaystyle{\rho^{m^-}=\lim_{n\to\infty}\frac{p_n^{m^-}}{p_n^{\Theta_0^-}}}$.  These estimates increase from $m=14$ to 18 and then decrease from $m= 18$ to 22.  On the other hand, if we consider instead the number of
 (-) juxtapositions that have a given $m^-$-size then:   the number that has size $m^-= 14, 16, 18, 20,  22$  is, respectively,  1, 10, 37, 60,  36. These numbers increase for $m=14$ to $20$ and then decrease for $m=20$ to 22, a slightly different trend than that observed for the proportion of polygons in each size class. Thus the numbers of $m^-$-size juxtapositions for $m=14$ to 22 do not determine the relative proportions of $m^-$-SAPs.

\begin{figure}[hhh]
\centering\psfrag{A}{\hspace{-.1in}$14_I$} \psfrag{B}{\hspace{-.1in}$16_I$} \psfrag{C}{\hspace{-.1in}$18_I$}
\psfrag{D}{\hspace{-.1in}$20_I$} \psfrag{E}{\hspace{-.1in}$22_I$} \psfrag{F}{$24$} \psfrag{G}{$26$}
\psfrag{H}{$28$} \psfrag{I}{$22$} \psfrag{J}{$32$} \psfrag{N}{$n_1$}
\psfrag{P}{\hspace{-.45in}$\hat{\rho}^{(m^-)}_{n_1,n_2}(\bar{\phi})$}
\includegraphics{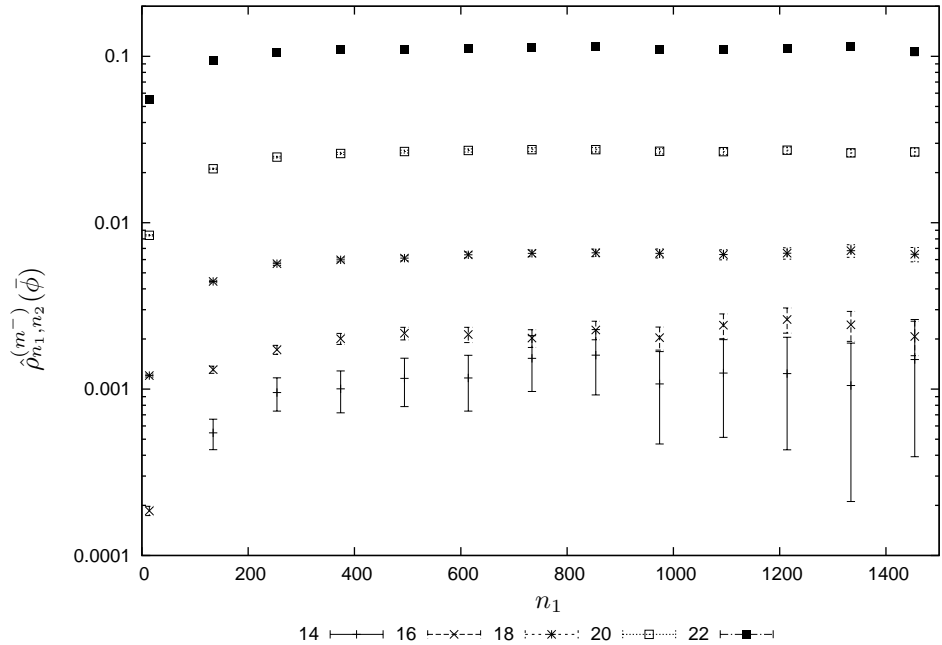}
\caption{The grouped-$n$ estimates for ${\rho}^{m^-}_{n_{1},n_{2}}(\bar{\phi})$, for (from bottom to top) $m\in\{14,16,18,20,22\}$ and $n_1\in\{14,134,254,...,1454\}$. The error bars represent 95\% confidence intervals.}
\label{juxtaposition_oneedge_perclass_trefoil}
\end{figure}

\begin{table}
\centering
\begin{tabular}{|c|c|c|}
  \hline
  $m$ & $\rho^{m^-}$& $\rho^{m^-}(\bar{\phi})$  \\\hline
  ${14}$ & $0.0222 \pm 0.0002$ & $0.0013 \pm 0.0002$ \\\hline
  ${16}$ & $0.1695 \pm 0.0005$ & $0.0027 \pm 0.0010$ \\\hline
 ${18}$ & $0.3867  \pm  0.0023$ & $0.0068  \pm  0.0003$\\\hline
  ${20}$ & $0.3165  \pm  0.0013$& $0.0269  \pm  0.0005$ \\\hline
   ${22}$ & $0.1032 \pm  0.0010$ & $0.1165 \pm  0.0074$\\
  \hline
\end{tabular}
\caption{In column 2, estimates of the limiting proportion $m^-$-SAPs along with 95\% margins of error. In column 3, estimates of the limiting knotting probability for $m^-$-SAPs along with 95\% margins of error.}
\label{table_95CI_compclassI}\label{table_95CI_propcompclassI}
\end{table}

Figure \ref{juxtaposition_oneedge_perclass_trefoil}
displays our estimates for ${\rho}^{m^-}_{n_{1},n_{2}}(\bar{\phi})$.  As $n_1$ goes to infinity, these estimates are expected to approach the limiting knotting  probability for $m^-$-SAPs, ${\rho}^{m^-}(\bar{\phi})$. \ Table~\ref{table_95CI_compclassI}
displays the estimates for these limiting knotting probabilities along with computed 95\% confidence intervals.

The juxtaposition ($Z^+$-mirror) associated with size-$14^-$ 
forms the tightest juxtaposition. Note that in figure \ref{juxtaposition_oneedge_perclass_trefoil},
 the grouped-$n$ estimates for ${\rho}^{14^-}_{n_{1},n_{2}}(\bar{\phi})$, for each $n_1\in\{14,134,254,...,1454\}$ are all very close to zero.  Consequently the limiting  knotting probability, $\rho^{14^-}(\bar{\phi})$,
will also be very close to zero; this is consistent with the observation in \cite{lmzc06}
 regarding their tightest (hooked) juxtaposition, that when starting with an unknot,  the after-virtual-strand-passage polygons
are essentially always unknotted. They also comment that their
after-virtual-strand-passage polygons are more knotted as the juxtaposition
becomes less tight.
\ Our results in table~\ref{table_95CI_compclassI} are consistent with
this trend since, statistically, they satisfy: \begin{equation}
{\rho}^{14^-}(\bar{\phi})<{\rho}^{16^-}(\bar{\phi})<{\rho}^{18^-}(\bar{\phi})<{\rho}^{20^-}(\bar{\phi})<{\rho}^{22^-}(\bar{\phi}).
\label{trend}
\end{equation}
This increasing trend with $m^-$ cannot be attributed to any trend in the proportions of polygons in classes $14^-$, $16^-$, $18^-$, $20^-$, and $22^-$ because, as previously noted, there is no strictly increasing trend in these proportions.

It should be pointed out, however, that the trend observed in equation (\ref{trend}) is an average property of the compactness classes.  Distinct juxtapositions having the same $m^-$-size can have quite different limiting knotting probabilities associated with them.  Furthermore, there exist distinct
juxtapositions, having different $m^-$-sizes, such that the associated limiting knotting probabilities follow the reverse trend to that of equation (\ref{trend}).  The observed trend of equation (\ref{trend}) is also
dependent on the choice of compactness measure; there are other possible choices for compactness measure, such as the dimensions of the smallest box containing a juxtaposition, where this trend is not observed.

\begin{figure}[hhh]
\centering\psfrag{14}{$14$} \psfrag{16}{$16$} \psfrag{18}{$18$}
\psfrag{20}{$20$} \psfrag{22}{$22$} \psfrag{F}{$24$} \psfrag{G}{$26$}
\psfrag{H}{$28$} \psfrag{N}{$n_1$}\psfrag{I}{$30$} \psfrag{J}{$32$} \psfrag{P}{\hspace{-.7in}$\hat{\rho}^{m}_{n_1,n_2}(\bar{\phi})$} \psfrag{N}{$n_1$}
\psfrag{M}[bl][l][1][90]{\hspace{-.7in}$\hat{\rho}^{m}_{n_1,n_2}(\bar{\phi}|s)$}
\includegraphics{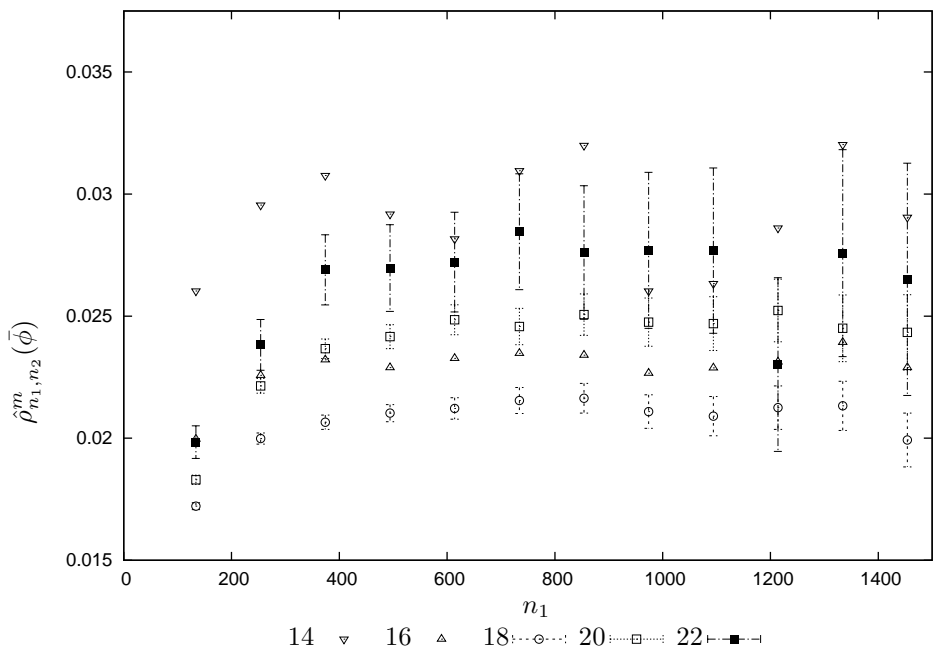}\caption{The grouped probability estimates of $\rho^{m}_{n_1,n_2}(\bar{\phi})$ for $m\in\{14 (\triangledown),16 (\vartriangle),18 (\circ),20 (\boxdot),22 (\blacksquare)\}$ and $n_1\in\{134,254,...,1454\}$. Only error bars (which represent 95\% confidence intervals) for compactness sizes-18, 20, and 22 are shown.   The estimated error bars for compactness sizes 14 and 16 are at least double those of the associated compactness size-20 error bar.}
\label{compactnessclasses_all}
\end{figure}

The compactness results just presented were based on taking into account crossing-sign.
If this is ignored, then amongst the 144 possible juxtapositions, the numbers with compactness size $m= 14, 16, 18, 20, 22$ respectively, are 2, 20, 68, 50, and 4.
Figure~\ref{compactnessclasses_all} displays the grouped probability estimates for $\rho^{m}(\bar{\phi})$
plotted versus  $n_1\geq 134$.  The corresponding estimates for $n_1=14$, which are an order of magnitude smaller than the others, are not plotted but given below:
$\rho^{14}_{14,134}(\bar{\phi})= 0.00502 \pm 0.00013$;
$\rho^{16}_{14,134}(\bar{\phi})= 0.00551 \pm 0.00006$;
$\rho^{18}_{14,134}(\bar{\phi})= 0.00613 \pm 0.00005$;
$\rho^{20}_{14,134}(\bar{\phi})= 0.00681  \pm 0.00007$;
$\rho^{22}_{14,134}(\bar{\phi})= 0.00706  \pm  0.00025$.
For this smallest choice of $n_1$, the point estimates (although very close to zero) do follow the general trend that the
more compact the polygon is around the strand passage site, the lower the associated probability of
knotting.  For the estimates in figure~\ref{compactnessclasses_all}, however, no such trend is present.
The most we can say is that compactness (when juxtaposition sign is ignored) does impact the associated limiting knotting probabilities.

\begin{figure}[ttt]
\centering\psfrag{259}{\tiny{$\widetilde{S^-}$}} \psfrag{252}{\hspace{-0.05in}\tiny{$S^{+}$}} \psfrag{511}{\hspace{-0.05in}\tiny{$\widetilde{Z^-}$}}
\psfrag{756}{\hspace{-0.05in}\tiny{$Z^{+}$}} \psfrag{Prob}{\hspace{-.1in}\tiny{$\hat{\rho}^{J^+}_{n_1,n_2}(\bar{\phi})$}} \psfrag{angle}{\small{$\alpha$}}
\subfigure[{}]{\includegraphics[scale=0.8]{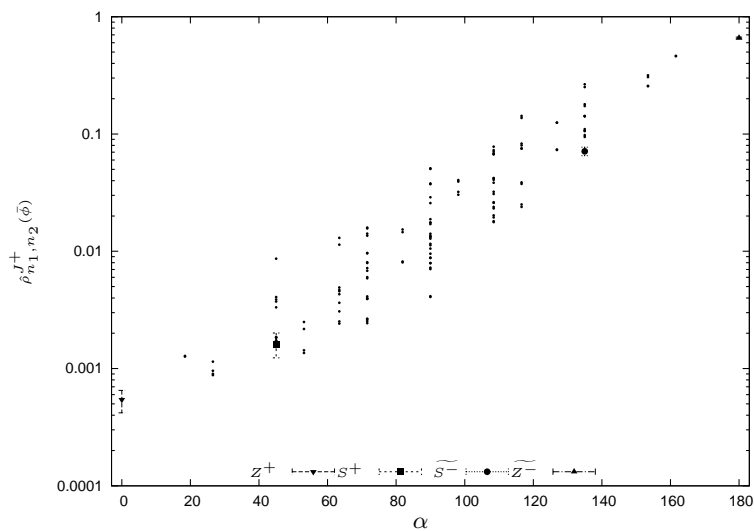}}\\\subfigure[{}]{\psfrag{Prob}{\hspace{-.1in}\tiny{$\hat{\rho}^{\alpha^+}_{n_1,n_2}(\bar{\phi})$}} \psfrag{N}{\small{${\alpha}$}}\psfrag{ave}{}
\includegraphics[scale=0.8]{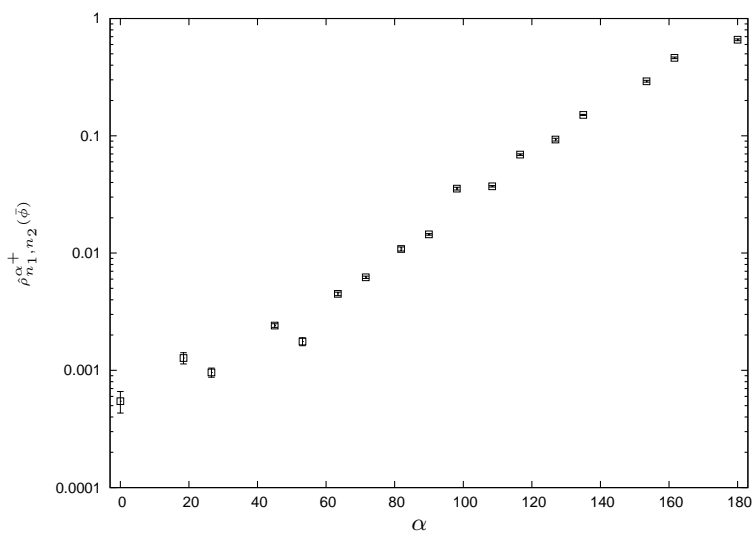}
}\caption{(a) The grouped probability estimates for the knotting probability for  ${J^+}$-SAPs whose lengths are between $n_1=134$ and $n_2=252$ edges inclusive, plotted versus opening angle $\alpha$ (degrees). The error bars presented  are estimated 95\% confidence intervals for $J^+\in\{\widetilde{Z^{-}}$, $\widetilde{S^{-}}$, $S^{+}$, $Z^{+}\}$;  the error bars  for all other juxtapositions are estimated to be smaller. (b) The grouped probabilities  for the angle-dependent knotting probabilities for  ${\alpha^+}$-SAPs whose lengths are between $n_1=134$ and $n_2=252$ edges inclusive, plotted versus opening angle. The error bars are estimated 95\% confidence intervals, with the majority being too small to appear clearly on the graph.}
\label{knotreductvsN1_94}
\end{figure}

Since crossing-sign plays a role and since the experimental
results (see \cite[Fig. 1]{neuman})  indicate that topo IV can have a preference for changing a (+) to
a (-) crossing, we focus on $\Theta^+$-SAPs for investigating the influence of the opening angle on knotting probability.  The results for $\Theta^+$-SAPs are 
obtained from our CMC sample of $\Theta^{-}$-SAPs by considering their
mirror image via $\widetilde{~~}$.  Recall that the opening angle was defined so that a juxtaposition $J^+$ had the same angle as $J^+$-mirror.
We first investigate the knotting probability for $\Theta_0^+$-SAPs  as a function of
the opening angle for all polygons with lengths $n=n_1=134$ to $n_2=252$ grouped together.  The knotting probability for each $J^+$, $J\in{\cal J}$ is
plotted versus opening angle in figure \ref{knotreductvsN1_94} (a).   The plot shows a positive
correlation between opening angle and  knotting probability.  This same trend was observed (plots are not shown here) for other choices of $n_1=14$ to $n_1=614$ and this was tested statistically.  The correlation coefficients for $n_1=14, 134, 254, 374, 494, 614$ are, respectively, $0.718$, $0.781$, $0.794$, $0.794$, $0.801$, $0.804$. For each case, we tested whether the true correlation associated with the data was zero versus the alternative that it was positive.  The $p$-values for each test are less than 0.001, hence we conclude that each correlation is positive. Thus as the opening angle increases from $0^\circ$ to $180^{\circ}$, on average, the probability of knotting increases, or equivalently the more acute the opening angle, the less likely (on average) the juxtaposition will knot an unknot.

\begin{figure}[hhh]
\centering\psfrag{180d}{\small{$180^\circ$}} \psfrag{0d}{\small{$0^\circ$}} \psfrag{53d}{\small{$53^\circ$}}
\psfrag{90d}{\small{$90^\circ$}} \psfrag{135d}{\small{$135^\circ$}} \psfrag{F}{$24$} \psfrag{G}{$26$}
\psfrag{H}{$28$} \psfrag{N}{$n_1$}\psfrag{I}{$30$} \psfrag{Prob}{$\hat{\rho}^{\alpha^+}_{n_1,n_2}(\bar{\phi})$} \psfrag{n1}{$n_1$}
\psfrag{M}[bl][l][1][90]{\hspace{-.7in}$\hat{\rho}^{J}_{n_1,n_2}(\bar{\phi}|s)$}
\includegraphics{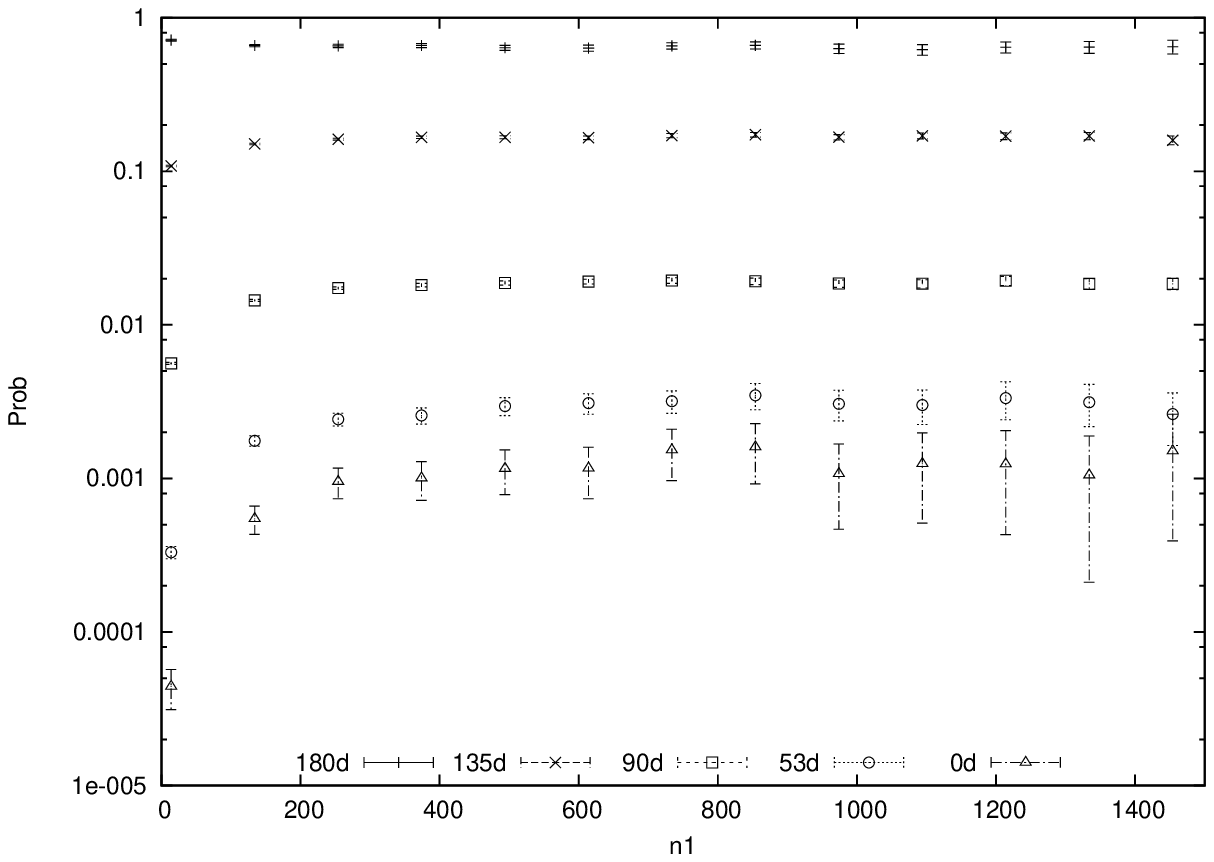}\caption{The grouped-$n$ estimates for ${\rho}^{\alpha^+}_{n_{1},n_{2}}(\bar{\phi})$, for (from bottom to top) $\alpha\in\{180,135,90,53,0\}$ and $n_1\in\{14,134,254,...,1454\}$. The error bars represent 95\% confidence intervals.}
\label{angle_134}
\end{figure}

To explore this further,  figure \ref{knotreductvsN1_94} (b)
presents the angle-dependent knotting probability obtained by grouping together the polygons whose juxtapositions have the same opening angle.   The figure supports that the knotting probability increases as opening angle increases, and appears very close to linear on the log scale (over the interval [0,180]).  To show that this trend continues as polygon length increases,  figure \ref{angle_134} displays the angle-dependent grouped-$n$  knotting probabilities for five angles (0$^\circ$, 53.13$^\circ$, 90$^\circ$, 135$^\circ$, 180$^\circ$) versus $n_1$.  The trend continues through all lengths.  We expect that
the knotting probabilities studied here will go to a constant as $n_1\to \infty$  and, for each angle, the point estimates appear to be approaching distinct angle-dependent limiting values.

\section{Summary and Discussion}
Topoisomerase enzymes are able to, through a local action (strand passage) on DNA, efficiently change the knot-type of a DNA molecule.   Just how the enzyme determines the local position within the DNA and to what extent randomness is involved are open questions.  Motivated by these questions, here we have presented a lattice polygon model for a local strand passage (cf. section~\ref{section2}), explored the asymptotic properties (as polygon length tends to infinity) associated with the model (cf. section~\ref{asymsec}), reviewed tools to simulate the model (cf. section~\ref{thetabfacfsection}) and to obtain statistical estimates (cf. sections~\ref{mlesec}-\ref{PROBABILITY_SECTION}), and then applied these tools to the simulated data to explore the asymptotic properties  (cf. sections~\ref{mlesec}-\ref{juxtasec}).

We prove that the number of $n$-edge unknotted polygons that contain a fixed structure grows at the same exponential rate $(\kappa_0)$ as the number of $n$-edge unknotted polygons.  We review a new maximum likelihood technique to estimate exponential growth rates and critical exponents using data generated from a composite Markov chain.  Using this technique, we obtain estimates for $\kappa_0$ which are consistent with other known estimates for $\kappa_0$.  We provide numerical evidence to support the conjecture that the limiting probabilities associated with different after-strand-passage properties  exist and lie strictly in $(0,1)$.  We also show that not only does the local geometry around the strand passage site influence the limiting probabilities of knotting, but combining this information with crossing-sign information has an even greater influence on  the limiting probabilities of knotting.
We further show, using two compactness measures that take the crossing-sign information into account, that as the local juxtaposition of a $\Theta_0$-SAP becomes more and more compact, the limiting probability of knotting associated with the compactness class decreases.

Others \cite{buck} have noted that the local geometry of the strand-passage site could play a role in the topoisomerase-DNA interaction.  Our work suggests that the crossing-sign at the strand passage site is also an important factor to consider, and this is consistent with experimental results which indicate that some type II topoisomerases exhibit a chirality bias.   Topoisomerase acts very locally on DNA, ie it acts on the DNA in the space occupied by the topoisomerase.  Consequently it acts within a finite volume, a very small volume when compared to the volume of the DNA itself.
Our model indicates that if the topoisomerase can take into account the crossing sign information at the strand passage site, then it can make a change in a small volume and preferentially knot an unknot (for our $Z^-$ juxtaposition, about 63\% of the time a strand passage  transforms the unknot into a knot) or preferentially leave it unknotted (for $Z^+$, only 0.13\% of the time does a strand passage result in a knot).

\ack

CES acknowledges support in the form of
Discovery and Equipment Grants from NSERC (Canada), resource allocations from Westgrid, and access to equipment funded by CFI. MLS acknowledges support in the form of  PMMB conference attendance funding, NSERC PGS A and B scholarships, and Univ. of Saskatchewan graduate funding.  The authors also acknowledge valuable input from D.W. Sumners, E.J. Janse van Rensburg, I. Darcy, S. Whittington, M. Vazquez, R. Scharein (for assistance with KnotPlot),  H.S. Chan,  E. L. Zechiedrich, Z. Liu and K. Neuman.

\appendix
\section{The CMC MLE method}

For the CMC MLE method, we assume that a sequence $\{{\boldsymbol{\hat{\omega}^{(t)}}},t=1,...,T\}$ of $M$-tuples ${\boldsymbol{\hat{\omega}}^{(t)}
=(\hat{\omega}_1^{(t}},...,\hat{\omega}_M^{(t)})$ of SAPs from a set ${\mathscr{S}}$ has been generated from a CMC Monte Carlo algorithm with
equilibrium distribution $\boldsymbol{\pi}(q,\boldsymbol{\beta})$; for our case, the SAPs are $\Theta^-$-SAPs and the distribution is given by  
\begin{equation}
{\boldsymbol{\pi}}(q,\boldsymbol{\beta}):=\{{\pi}_{\boldsymbol{\omega}}(q,\boldsymbol{z}),\boldsymbol{\omega}\in\mathscr{S}^M\},
\label{proddist}
\end{equation}
where $\boldsymbol{\beta}=({\beta_1},{\beta_2},...,{\beta_M})$ with ${\beta_i}<{-\kappa_0}$ for each $i$,  $\boldsymbol{z}=(e^{\beta_1},e^{\beta_2},...,e^{\beta_M})$, and ${\pi}_{\boldsymbol{\omega}}(q,\boldsymbol{z})$ is as in equation (\ref{CMCpmfb}).
Let $s_n$ denote the number of $n$-edge polygons in $\mathscr{S}$.
It is also assumed that for any of the polygon subsets of interest, $*$-SAPs in our case with $*\in\{\Theta^-,\Theta_0^-,\Theta_0^-(K);K\in{\cal{K}}\}$,  that the number of $n$-edge
polygons in the subset, $s_n^*$, satisfies
for all
$n\geq N_{\min}^{\ast}$, 
\begin{equation}
s_{n}^{\ast}=A_{\ast}(n+h_{\ast})^{\alpha_{\ast}}e^{n\kappa_{0}}. \label{assump2_ini}
\end{equation}
Let 
$s_n^{\bar{\ast}}=s_n-s_n^{\ast}$; analogous assumptions are made about the asymptotic behaviour of $s_n^{\bar{\ast}}$.
Note that the ``$h$'' in this general asymptotic form is added to accommodate for
some of the effects of the unknown $o(1)$ term in equation (\ref{asymform}).
Finally, we assume that the probability, denoted by $Q_{N_{\min}^*}(\beta_i)$, that a chain $i$ polygon's length is less than $N_{\min}^*$ under distribution ${\boldsymbol{\pi}}(q,{\boldsymbol{\beta}})$ is an unknown parameter for each $i=1,...,M$.
With these model assumptions, the goal of the method is then to obtain maximum likelihood estimates for $\alpha_*$ and $\kappa_0$ and the other unknown parameters
using the CMC Monte Carlo data.  To do this, given a specific subset or ``property'' $*$, a log-likelihood function needs to be defined.
The relevant log-likelihood function is defined below.

Note that the model's asymptotic form (from equation (\ref{assump2_ini}))
depends only on polygon lengths and the property *. Furthermore, the model applies only
to sufficiently large polygon lengths.  At the same time, very large polygons are rare events in a Markov chain generated from the $\Theta$-BFACF algorithm.
Thus we concentrate on polygon observations in a restricted polygon-length interval where it is expected that the model applies and that there is sufficient data for reliable estimates.
For this purpose, given a property $\ast$ and fixed even positive integers $N$ and $N'$ (the boundaries of the polygon length interval of interest) such that $14<N<N'$, we define the following four indicator functions for $\omega \in \mathscr{S}$:
\begin{equation}
\psi_{\ast}(\omega):=\left\{ \begin{array}{ll}
1, & \mbox{\rm if }\omega ~\mbox{\rm has property \ensuremath{\ast}}\\
0, & \mbox{\rm otherwise}\end{array}\right.,\end{equation}
 and for any even positive integer $n$, define\begin{equation}
I_{\left\langle 1\right\rangle }(n):=\left\{ \begin{array}{ll}
1, & \mbox{\rm if }0\leq n<N\\
0, & \mbox{\rm otherwise}\end{array}\right.,\end{equation}
 \begin{equation}
I_{\left\langle 2\right\rangle }(n):=\left\{ \begin{array}{ll}
1, & \mbox{\rm if }N\leq n\leq N'\\
0, & \mbox{\rm otherwise}\end{array}\right.,\mbox{\rm }\end{equation}
 and\begin{equation}
I_{\left\langle 3\right\rangle }(n):=\left\{ \begin{array}{ll}
1, & \mbox{\rm if }n>N'\\
0, & \mbox{\rm otherwise}\end{array}\right..\end{equation}
Next, let $|\omega|$ denote the length of a polygon $\omega\in{\mathscr{S}}$.
Our interest is in two functions, defined on $\mathscr{S}$, of these indicator functions:
\begin{equation}
X(\omega)=I_{\left\langle 2\right\rangle}(|\omega|)|\omega|+I_{\left\langle 3\right\rangle}(|\omega|)(N'+1),
\end{equation}
which, within the restricted interval, keeps track of the actual polygon lengths, while, outside the restricted interval, it just keeps track of whether the length is above or below the interval boundaries; and
\begin{equation}
X_*(\omega)=I_{\left\langle 2\right\rangle}(|\omega|)\psi_*(\omega),
\end{equation}
which keeps track of whether a polygon in the restricted length interval has the given property or not.  The possible values for the pair $(X(\omega),X_*(\omega))$ are given by $(n,\eta)\in S(N,N'):=\{(0,0),(N'+1,0)\}\cup\{\{N,N+1,...,N'\}\times \{0,1\}\}$.  Thus, given $N$ and $N'$, 
from the generated CMC Monte Carlo polygon data we can obtain a sequence of  $M$-tuples of ordered pairs $(X(\hat{\omega}_i^{(t)}),X_*(\hat{\omega}_i^{(t)}))=(\hat{n}_i^{(t)},\hat{\delta}_i^{(t)})\in S(N,N')$, for $i=1,...,M$, $t=1,...,T$.

For $N=N_{\min}^*$ and $N'=N_{\max}^*$, consider any $(\boldsymbol{n}_T,\boldsymbol{\delta}_T)$ that is a sequence  of $T$ M-tuples of pairs $(n^{(t)}_i,\delta^{(t)}_i)\in S(N,N')$, $i=1,...,M$, $t=1,...,T$.  The
model log-likelihood, ${{\ell}_{T}^{\prime}}$, for this sequence as an outcome from the CMC Monte Carlo algorithm can be written in terms of the 
$(M+6)$ unknown parameters ($\kappa_{0},$ $\alpha_{\ast},$ $h_{\ast},$
$\alpha_{\overline{\ast}},$ $h_{\overline{\ast}},$  $A=\frac{A_{\ast}}{A_{\overline{\ast}}}$ and
 $\tilde{Q}(\beta_{i})=1-Q_N(\beta_i)),$ for $i\in\{1,2,...,M\},$
as follows \cite{s09}:
\begin{eqnarray}
{\ell}_{T}^{\prime} & :=T^{\prime}\sum_{i=1}^{M}\left[\left\langle I_{\left\langle 2\right\rangle }(n_{i})\log w(n_{i})\right\rangle _{T}+(\kappa_{0}+\beta_{i})\left\langle I_{\left\langle 2\right\rangle }(n_{i})n_{i}\right\rangle _{T}\right]\nonumber \\
 & +T^{\prime}\sum_{i=1}^{M}\left[\alpha_{\ast}\left\langle \delta_{i}\log(n_{i}+h_{\ast})\right\rangle _{T}\right]\nonumber \\
 & +T^{\prime}\sum_{i=1}^{M}\alpha_{\overline{\ast}}\left\langle \left[ I_{\left\langle 2\right\rangle }(n_{i})-\delta_{i}\right]\log(n_{i}+h_{\overline{\ast}})\right\rangle _{T}\nonumber \\
 & +T^{\prime}\sum_{i=1}^{M}\left(\left\langle \delta_{i}\right\rangle _{T}\log A+\left\langle I_{\left\langle 1\right\rangle }(n_{i})\right\rangle _{T}\log\left[1-\tilde{Q}\left(\beta_{i}\right)\right]\right)\nonumber \\
 & +T^{\prime}\sum_{i=1}^{M}\left\langle I_{\left\langle 3\right\rangle }(n_{i})\right\rangle _{T}\log\left[Q_{\left\langle 3\right\rangle }^{\overline{\ast}}\left(\beta_{i}\right)+AQ_{\left\langle 3\right\rangle }^{\ast}\left(\beta_{i}\right)\right]\nonumber \\
 & +T^{\prime}\sum_{i=1}^{M}\left\langle I_{\left\langle 2,3\right\rangle }(n_{i})\right\rangle _{T}\left[\log\tilde{Q}\left(\beta_{i}\right)-\log\left[Q_{\left\langle 2,3\right\rangle }^{\overline{\ast}}\left(\beta_{i}\right)+AQ_{\left\langle 2,3\right\rangle }^{\ast}\left(\beta_{i}\right)\right]\right], \label{loglikelihood_modified}\end{eqnarray}
 where  $I_{\left\langle 2,3\right\rangle }=I_{\left\langle 2\right\rangle }+I_{\left\langle 3\right\rangle }$, 
 \begin{equation}
Q_{\left\langle 3\right\rangle }^{\bullet}\left(\beta\right):={\displaystyle \sum\limits _{n>N_{\max}^{\ast}}}w(n)\left(n+h_{\bullet}\right)^{\alpha_{\bullet}}e^{\left(\kappa_{0}+\beta\right)n},\end{equation}
\begin{equation}
Q_{\left\langle 2,3\right\rangle }^{\bullet}\left(\beta\right):={\displaystyle \sum\limits _{j\geq N_{\min}^{\ast}}}w(j)\left(j+h_{\bullet}\right)^{\alpha_{\bullet}}e^{\left(\kappa_{0}+\beta\right)j},
\end{equation}
$\bullet\in\{\ast,\overline{\ast}\}$,
$w(n)=(n-6)n^q$
and, for any function $g$ defined on $S(N_{\min}^{\ast},N_{\max}^{\ast})$,
\begin{equation}
\left\langle g(n_{i},\delta_{i})\right\rangle _{T}=\frac{\sum_{t=1}^{T}g(n_{i}^{(t)},\delta_{i}^{(t)})}{T}.\label{sample_average}
\end{equation}
 Note that the sample averages in ${{\ell}_{T}^{\prime}}$
are based on all $T$ sample data points and are given by equation
(\ref{sample_average}). \ Also note that ${{\ell}_{T}^{\prime}}$ is only a function of the $(M+6)$ parameters: $\kappa_{0},$ $\alpha_{\ast},$ $h_{\ast},$
$\alpha_{\overline{\ast}},$ $h_{\overline{\ast}},$  $A$ and
 $\tilde{Q}(\beta_{i}),$ for $i\in\{1,2,...,M\}.$
The factor $T'$ in front accomodates for the fact that the generated
data is not necessarily independent.
Following the Berretti-Sokal Method \cite{bs85}, to compensate
for the lack of independence in the sample data, the log-likelihood function obtained under the assumption of independence can be  rescaled according to the number of ``essentially
independent'' data points, $T^{\prime}$, by multiplying it by $T^{\prime}/T$.
The result in this case, is the log-likelihood function defined above.

To obtain the MLEs for the parameters $\kappa_{0},$ $\alpha_{\ast},$ $h_{\ast},$
$\alpha_{\overline{\ast}},$ $h_{\overline{\ast}},$  $A$ and
 $\tilde{Q}(\beta_{i}),$ for $i\in\{1,2,...,M\},$ in the log-likelihood $\ell'_T$: $\ell'_T$ is differentiated with respect to each of these $M+6$ parameters;  each of the resulting partial derivatives is set to zero; and finally the resulting system of equations is solved simultaneously to obtain the MLEs.  For $1 \leq i \leq M$, setting $\frac
{\partial{{\ell}_{T}^{\prime}}}{\partial{Q}\left(  \beta
_{i}\right)  }=0$ and then solving for $\tilde{Q}\left(  \beta_{i}\right)  $
yields the following MLEs for $\tilde{Q}\left(  \beta_{i}\right)  :$
\begin{equation}
\tilde{Q}\left(  \beta_{i}\right)  =\left\langle I_{\left\langle
2,3\right\rangle }(n_{i})\right\rangle _{T},\mbox{\rm for }i\in\{1,2,...,M\}.
\label{MLE_Qtildai}
\end{equation}
In order to obtain MLEs for the remaining six parameters ($\kappa_{0},$ $\alpha_{\ast},$ $h_{\ast},$
$\alpha_{\overline{\ast}},$ $h_{\overline{\ast}},$  and $A$), the MLEs from
equation~(\ref{MLE_Qtildai}) are substituted into the system of  equations to yield a new system of six equations.   The new system is solved numerically for the MLEs using the Newton-Raphson Method.

In practice, the simulated data is used to select an appropriate choice
for the boundaries, $N$ and $N'$, of the restricted polygon length interval.
For a given property $\ast$, we first estimate, based on the statistical ``reliability'' of the generated polygon length data, a value for  $N'=N_{\max}^{\ast}$.
Then,
a choice $N=N_{\min}^{\ast}$ is obtained as follows.  Given any $N$, let $\hat{\kappa}_0(N)$ and $\hat{\alpha}_{\ast}(N)$ be CMC MLE estimates for $\kappa_0$ and
$\alpha_{\ast}$, respectively. Then $N_{\min}^{\ast}$ is estimated to be the first value of $N$ in $\{14,16,18, ...\}$ for
which for all $m$ such that $N\leq m < N_{\max}^{\ast}$, $|\hat{\kappa}_0(m)- \hat{\kappa}_0(m+2)|<0.000001$ and
$|\hat{\alpha}_{\ast}(m)- \hat{\alpha}_{\ast}(m+2)|<0.0001$.
In other words, we choose $N_{\min}^{\ast}$ to be the value for which the estimates $\hat{\kappa}_0(N)$ and
$\hat{\alpha}_{\ast}(N)$ are essentially constant for all $N\geq {N}_{\min}^{\ast}$.  This is akin to the so-called ``flatness'' region discussed in \cite{bs85}.  Full details of the methods for choosing $N_{\max}^{\ast}$
and $N_{\min}^{\ast}$ are given in \cite{s09}.

\end{document}